\documentclass[twocolumn,latexsym,showpacs,verbatim,preprintnumbers,amsmath,amssymb,nofootinbib,superscriptaddress]{revtex4-2}
\usepackage{graphicx,natbib,float}
\usepackage{xcolor}
\usepackage{dcolumn}
\usepackage{bm}
\usepackage{upgreek}
\usepackage{romannum}
\usepackage{footmisc}

\begin{document}
\title{A new potential method for the $X_\mathrm{max}$ measurement of extensive air showers based on backtracking radio signals}
\author{V.B.~Jhansi}
\email{vuta@fzu.cz}
\affiliation{Department of Physics, Khalifa University of Science and Technology, P.O. Box 127788, Abu Dhabi, United Arab Emirates}
\affiliation{Now at: Department of Astroparticle Physics, Institute of Physics of the Czech Academy of Sciences, 18200 Prague, Czech Republic.}

\author{S.~Thoudam}
\email{satyendra.thoudam@ku.ac.ae}
\affiliation{Department of Physics, Khalifa University of Science and Technology, P.O. Box 127788, Abu Dhabi, United Arab Emirates}

\author{S.~Buitink}
\affiliation{Astrophysical Institute, Vrije Universiteit Brussel, Pleinlaan 2, 1050 Brussels, Belgium}
\affiliation{ Interuniversity Institute for High-Energy, Vrije Universiteit Brussel,  Pleinlaan 2, 1050 Brussels, Belgium}

\author{A.~Corstanje}
\affiliation{ Interuniversity Institute for High-Energy, Vrije Universiteit Brussel,  Pleinlaan 2, 1050 Brussels, Belgium}

\author{M.~Desmet}
\affiliation{Astrophysical Institute, Vrije Universiteit Brussel, Pleinlaan 2, 1050 Brussels, Belgium}

\author{J.~R.~H\"orandel}
\affiliation{Department of Astrophysics / IMAPP, Radboud University Nijmegen,  P. O. Box 9010, 6500 GL, Nijmegen, The Netherlands}
\affiliation{NIKHEF, Science Park Amsterdam, 1098 XG Amsterdam, The Netherlands}

\author{T.~Huege}
\affiliation{Institut f\"{u}r Astroteilchenphysik (IAP), Karlsruhe Institute of Technology (KIT), P. O. Box 3640, 76021, Karlsruhe, Germany}
\affiliation{Interuniversity Institute for High-Energy, Vrije Universiteit Brussel,  Pleinlaan 2, 1050 Brussels, Belgium}

\author{K.~Mulrey}
\affiliation{Department of Astrophysics / IMAPP, Radboud University Nijmegen,  P. O. Box 9010, 6500 GL, Nijmegen, The Netherlands}
\affiliation{NIKHEF, Science Park Amsterdam, 1098 XG Amsterdam, The Netherlands}

\author{O.~Scholten}
\affiliation{Interuniversity Institute for High-Energy, Vrije Universiteit Brussel, Pleinlaan 2, 1050 Brussels, Belgium}
\affiliation{ University of Groningen, Kapteyn Astronomical Institute, Groningen, 9747 AD, Netherlands}

\date{\today}
 
\begin{abstract}
{Measurements of cosmic-ray composition based on air-shower measurements rely mostly on the determination of the position of the shower maximum ($X_\mathrm{max}$). One efficient technique is to image the development of the air shower using fluorescence telescopes. An alternative technique that has made significant advances in the recent years is to measure the radio emission from air shower. Common methods for $X_\mathrm{max}$ determination in the radio detection technique include fitting a two-dimensional radio intensity footprint at the ground with Monte-Carlo simulated showers which is computationally quite expensive, and others that are based on parameterizations obtained from simulations. In this paper, we present a new method which is computationally extremely efficient and has the potential to reconstruct $X_{\rm max}$ with minimal input from simulations. The method involves geometrical reconstruction of radio emission profile of air showers along the shower axis by backtracking radio signals recorded by an array of antennas at the ground. On implementing the method on simulated cosmic-ray proton and iron showers in the energy range of $\rm 10^{17}-10^{18}\,eV$, we find a strong correlation between the radio emission profile obtained with the method in the $20-80$~MHz frequency range and the shower longitudinal profile, implying a new potential way of measuring $X_\mathrm{max}$ using radio signals.}
\end{abstract}

\maketitle
\section{Introduction}
Cosmic rays (CRs) were discovered by Victor Hess in 1912 after the observation of an increase in the rate of discharge of electroscopes with altitude. 
Even after more than a century since their discovery, the origin of CRs remains one of the fundamental problems in Astrophysics. 
Cosmic rays are predominantly charged particles consisting mainly of protons, helium nuclei and a very small fraction of heavier nuclei. They have been measured up to energies exceeding $\rm 10^{20}\,eV$ with a spectrum that follows, in general, a power law behavior of the form $I(E)\propto E^{-\gamma}$. The spectral index $\gamma$ changes from 2.7 to 3.1 at around energy $\rm E\sim 10^{15}\,eV$, the so called  ``knee'' in the CR spectrum, and becomes flatter to $\gamma\sim 2.7$ at $\rm E\sim 10^{18}\,eV$, which is referred to as the ``ankle'' in the CR spectrum. Another feature of the spectrum which is less prominent is the slight steepening at $\sim10^{17}$~eV, usually referred to as the ``second knee'' \cite{tanabashi2018review}. Numerous measurements of the CR energy spectrum and their elemental composition by different experiments, together with the observations of non-thermal emission from potential sources of high-energy particles in the Universe, have pushed forward our understanding of the  origin of CRs (see Ref. \cite{tatischeff2021origin} for a recent review), yet the exact nature of their sources still remain an open question. The lowest energy CRs ($\rm \lesssim 100\,MeV$) most likely originate from the Sun. Beyond this energy, CRs are considered to be mainly produced by supernova remnants in the Galaxy, with those in the highest energy region above $\sim 10^{18}$~eV coming from external galaxies \cite{blasi2013origin,biermann2010origin,hillas2005can,thoudam2016cosmic,zirakashvili2006cosmic}. The maximum energy up to which the Galactic CRs extend is not clearly understood. But, it is widely believed that the transition from the galactic-to-extragalactic CRs occurs in the energy region of $10^{17}-10^{18}$\,eV (see e.g., Refs. \cite{thoudam2016cosmic} and references therein), which is also the energy region where the LOFAR measures CRs \cite{buitink2016large}.

Theoretical models predict a composition of CRs in the transition energy region that changes from heavy to light elements with energy. However, recent measurements from experiments such as LOFAR and the Pierre-Auger Observatory indicate a composition between $\sim 10^{17}$ and $10^{18}$~eV which largely consists of light and intermediate nuclei with weak or negligible iron fraction, in contrast to most common assumptions \cite{aab2014depth,buitink2016large}. These observations might indicate the presence of an additional component of Galactic CRs such as reacceleration of CRs at the Galactic wind termination shocks, CRs originating from supernova remnants associated with Wolf-Rayet stars \cite{thoudam2016cosmic} and CRs accelerated by star clusters in the Galaxy \cite{bhadra2024between}. All these different models predict a different CR composition in the transition energy region. A precise measurement of composition in the transition region is thus crucial for a better understanding of the origin of CRs as well as their propagation in the Galaxy and extragalactic medium.

High-energy CRs produce extensive air showers (EAS) upon interaction with the Earth's atmosphere. Ground-based experiments such as KASCADE \cite{apel2013ankle} and ICETOP \cite{aartsen2013measurement} measure CRs through the direct detection of secondary particles contained in the EAS. These experiments measure the CR composition based on the observed electron-to-muon ratio of the EAS. An alternative method is to measure the electromagnetic radiation from the EAS, and determine the mass (type) of the CR particle based on the reconstructed longitudinal profile of the shower or the nature of the shower footprint at the ground.

The electromagnetic component of an EAS (which comprises mostly of electrons, positrons and gammas) reaches maximum at a certain atmospheric depth, the so-called depth of the shower maximum ($X_\mathrm{max}$). On average, lighter nuclei such as proton and helium have larger $X_\mathrm{max}$ values than heavier nuclei such as iron at the same given energy. Measurements of CR composition with ground-based experiments primarily rely on the measurements of $X_\mathrm{max}$. The  Pierre-Auger Observatory and Telescope Array experiments use fluorescence telescopes to measure the longitudinal profile of the shower by recording the fluorescent emission from the nitrogen molecules excited by EAS particles in the atmosphere \cite{abraham2010fluorescence,tokuno2012new}. Although fluorescence telescopes can give an $X_\mathrm{max}$ resolution of $\rm \sim 20\,g\,cm^{-2}$ above $\rm\sim 10^{18}$~eV, they suffer limitations from high construction cost and short duty cycle (less than about 20\%) as they can be operated only during dark moonless and clear nights. The Tunka-133 experiment implements an alternative technique that is based on the detection of Cherenkov emission from EAS, and obtain an $X_\mathrm{max}$ resolution of $\rm \sim 28\,g\,cm^{-2}$ above $\rm\sim 10^{16.5}$~eV \cite{prosin2014tunka}. But, this technique also suffers from small duty cycle. On the other hand, the radio measurement technique which is based on the detection of radio emission from EAS offers low costs, an almost $100\%$ duty cycle and an $X_\mathrm{max}$ precision comparable to that of the fluorescence telescopes \cite{buitink2014method}.

  
  One of the first methods of $X_\mathrm{max}$ measurement using radio data, as adopted by the LOPES experiment, is to initially parameterize the dependence of the slope of the  radio shower front on $X_\mathrm{max}$ through  simulations, and then determine $X_\mathrm{max}$ by applying the parameterization on the measured data \cite{apel2014wavefront}. An alternative method involves parameterization of the slope of the radio lateral distribution \cite{apel2014reconstruction}. Using these methods, the LOPES experiment obtained an $ X_\mathrm{max}$ resolution of $\rm \sim 100\,g\,cm^{-2}$.  In more sophisticated antenna arrays such as LOFAR and Tunka-Rex, accurate parameterization of the radio lateral distribution gave an improvement in the $X_\mathrm{max}$ resolution to  $\rm \sim 40\,g\,cm^{-2}$ above $\rm\sim 10^{17}$~eV \cite{nelles2015radio,bezyazeekov2016radio}. The current best method adopted by the LOFAR experiment involves comparison of the measured radio footprint to that of an ensemble of simulated showers to derive the $X_\mathrm{max}$ that best fits the data. This method yields an $X_\mathrm{max}$ resolution of $\sim \rm 17\,g\,cm^{-2}$ for LOFAR \cite{buitink2014method}, which is close to that measured with the fluorescence telescopes and also comparable with the resolution from the Auger Engineering Radio Array (AERA) of the Pierre Auger Observatory  \cite{abdul2024demonstrating,abdul2024radio}. However, this method is computationally quite expensive, making it not so efficient to handle a large volume of data. This limitation will be even more severe when applied to the upcoming SKA telescope which will have a much denser antenna population than LOFAR \cite{corstanje2023simulations}.
  
  In interferometric techniques, $X_\mathrm{max}$ is obtained by parameterizing the depth of the maximum coherent power where the coherent signal at a particular point is obtained by summing over the time-shifted signal traces of the antennas at the ground level.
  It has been shown that this method can reach an $X_{\rm max}$ resolution of $\rm \sim 10\,g\,cm^{-2}$ for a highly idealized simulation set-up \cite{schoorlemmer2021radio}. An improvement of the interferometric method which overcomes the limitations due to the finite aperture effects of the antenna array has been recently presented in Ref. \cite{scholten2024press}.
  
The main drawback of the radio detection methods described above is that, unlike the fluorescence technique, they cannot determine $X_\mathrm{max}$ purely from the measured data, independent of the simulation. Therefore, the $X_\mathrm{max}$ values obtained can be affected by the simulation settings.
In this paper, we present a novel and promising method which has the potential to reconstruct $X_\mathrm{max}$ with minimal inputs from simulation, and at the same time, computationally quite efficient. The method is based on the geometrical reconstruction of radio emission profile of EAS by backtracking radio signals observed by an array of antennas at the ground to their emission points. The reconstruction uses information about the shape of the radio wavefront and the lateral distribution of the radio signals at the ground. 
The paper is organized as follows. The mechanism of radio emission from EAS is discussed in Section-$\rm \Romannum{2}$. In Section-$\rm \Romannum{3}$, the geometry of the antenna array and the simulation set-up adopted in the present work are described. In Section-$\rm \Romannum{4}$, we describe the new potential method for the $X_\mathrm{max}$ reconstruction, the various steps followed for the reconstruction of the arrival direction and the core position of the EAS, and the reconstruction of the radio emission profile. 
  Section-$\rm \Romannum{5}$ presents a comparison of the reconstructed radio emission profile with the shower longitudinal profile, Section-$\rm \Romannum{6}$ discusses the results, and Section-$\rm \Romannum{7}$ presents the conclusion of this work.

\section{ Radio emission from EAS}
Radio emission from EAS is primarily attributed to the mechanism of geomagnetic emission. The electrons and positrons in the EAS drift apart as they interact with the geomagnetic field, producing a transverse current in the shower front. The transverse current varies in proportion to the total charge of the electrons and positrons in the EAS, and thereby radio emission is produced. The electric fields produced via this mechanism are polarized along the direction of the cross-product of the propagation vector of the EAS $\overrightarrow{\rm V}$ and the geomagnetic field $\overrightarrow{\rm B}$ \cite{schroder2017radio}. Geomagnetic emission contributes $\sim\,80\%$ - $95\%$ of the total pulse amplitude, depending on the local magnetic field strength, and multiple geometrical factors such as the observer position, zenith angle and the angle to the geomagnetic field \cite{buitink2014method}. An additional contribution to the radio emission results from the changing current along the shower axis, produced as a result of the excess number of electrons over positrons at the EAS front which is generated due to the knocking-off of electrons from the air molecules by particles in the EAS and the annihilation of positrons with the atmospheric electrons. The electric fields produced by this mechanism of charge excess emission point radially inwards towards the shower axis \cite{schroder2017radio}.

Radio emission from the two different mechanisms described above become coherent and amplified at wavelengths  that are comparable to the thickness of the shower front giving an emission which is typically observed in the frequency band of tens of MHz. The two radiation components undergo superposition, and this  gives rise to a rotationally asymmetric distribution of radio intensity at the ground due to the different polarization behavior of the components. The radio intensity pattern is further complicated by Cherenkov-like effects, owing to the increase of the refractive index of air with atmospheric depth.
This effect leads to coherent radio emission up to $\sim$ GHz frequencies with the coherence effect mostly limited to a certain angle with respect to the shower axis \cite{nelles2015measuring}.

\section{Simulation}

We use pre-existing simulated showers from the LOFAR repository. The showers were simulated using the CORSIKA (version 7.7410) air-shower simulation package \cite{heck1998corsika} and the CoREAS extension for radio simulation \cite{huege2013simulating}, considering QGSJET-II-04 and UrQMD models for the hadronic interaction. CoREAS simulations involve calculation of radio emission from first principles in which the radio emission from each individual electron and positron in the EAS is obtained based on the the end-point formalism \cite{huege2013simulating}. The pre-existing radio simulations were performed for a set of 160 antennas placed at an altitude of $\rm 7.6\,m$ above sea level which corresponds to the elevation of the LOFAR site. The antennas are distributed in such a way that they form a star-shape radial grid when projected onto the shower plane. The grid extends up to $\rm 500\,m $ in the shower plane with antenna separation of $\rm 20\,m$.

The atmospheric profile for the CORSIKA simulations were set according to the GDAS model, suited for the local atmospheric conditions at LOFAR \cite{mitra2020reconstructing,will2014global}. To save computation time, a thinning level of $10^{-6}$ was set with an optimized weight factor suitable for the radio emission. The signals were generated in a time window of 480\,ns, with a resolution of $\rm 0.1\,ns$. 

 For the present work, we use 400 antennas placed at the LOFAR elevation and distributed in a square array of size $\rm 400\,m\times400\,m$ with an antenna spacing of $\rm 20\,m$. The signals at each of the 400 antennas in this array are obtained by interpolating  the signals of the star-shaped antenna grid described above. The interpolation is performed using a high-order interpolation algorithm, to yield the polar components of the electric field, $E_{\theta}$ and $ E_{\phi}$, in the frequency range of 20-80 MHz which typically corresponds to the frequency range of the LOFAR low-band antenna. This algorithm is based on the Fourier representation  of signal around circles in the shower plane, and then using cubic splines to interpolate the Fourier coefficients along the radial direction. The details of the method can be found in Ref. \cite{corstanje2023high}. The interpolation does not introduce any significant artifacts in the shower footprint and the shape of the radio wavefront at the ground.
 
 For the analysis presented in this paper, we use a total of $\rm 24\times 10^{3}$ proton EAS and $13\times 10^{3}$ iron EAS in the  zenith angle range of $\rm 0^{\circ}-40^{\circ}$, and an energy range of $\rm 10^{17}-10^{18}\,eV$. These EAS have their shower core positions located at the centre of the square array.

\section{The new potential $X_\mathrm{max}$ reconstruction method}
\label{method}
It has been demonstrated in earlier experiments that the radio footprint pattern observed at the ground is sensitive to the longitudinal profile of the EAS. Therefore, it should be possible to reconstruct the longitudinal profile using the radio footprint observed at the ground. The new method presented here involves backtracking of the radio signals recorded by the antennas to their emission points along the shower axis considering that the signals travel perpendicular to the radio wavefront.

\begin{figure}
\includegraphics[width=1\columnwidth]{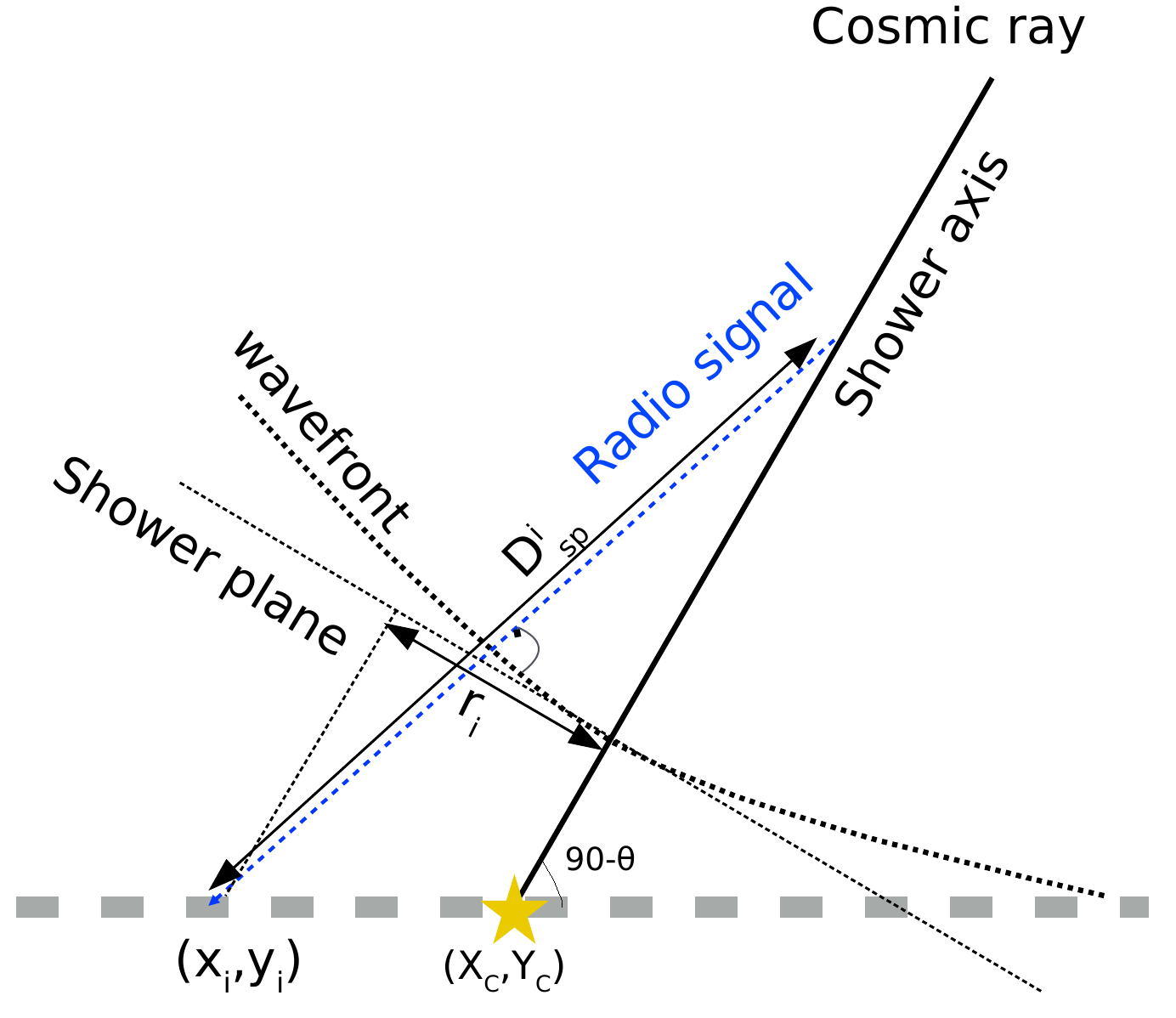}
\caption{A schematic of the reconstruction method  used for obtaining $D_{SP}$.
For an antenna located  at $(X_{i},Y_{i})$ in the array, $D_{SP}^{i}$ is obtained by assuming that the signal travels perpendicular to the non-planar wavefront. The  distance of the $\rm i^{th}$ antenna from the shower core $(X_C,Y_C)$ is projected on to the shower plane to obtain $r_{i}$. The zenith angle of the EAS is $\rm \theta$.}
\label{showerfront}
\end{figure}

At each antenna, a straight line, hereafter ``ray'', is constructed perpendicular to the radio wavefront. The ray is backtracked to determine the point of intersection of the ray with the shower axis. The shower axis is represented by the line drawn from the position of the shower core towards the arrival direction of the shower. The shower core position and the arrival direction are reconstructed as described below in Section \ref{subsection:core}. The point of intersection thus obtained is considered as the ``effective'' source point of the signal received by the antenna. The radio emission profile of the EAS is then obtained by constructing a distribution of the effective source points along the shower axis where each point is scaled by the fluence\footnote{Integration of the instantaneous Poynting vector over a symmetric window around the pulse maximum} received by the antenna and the square of the distance of the source point from the antenna position at the ground $D_{\rm sp}$ as shown in Fig.\ref{showerfront}. In the following, we demonstrate that the radio emission profile obtained with this method shows a strong correlation with the longitudinal profile of EAS, indicating that the peak position of the reconstructed radio profile can be a good measure of the position of the shower maximum.

As the ultimate goal is to apply the method to real experimental data, we follow an analysis procedure that would be applicable to measured data. As in real data, output from CoREAS simulation gives the total radio signal at each antenna. First, as described in Section \ref{subsection:core}, the arrival direction and the shower core position are determined based on the total signals received by the antennas. These shower parameters are then used to separate the geomagnetic signal from the charge-excess component  as described in Section \ref{subsection:sep}. For the reconstruction of the radio emission profile, we use the dominant geomagnetic component of the radio signal. This is described in Section \ref{reco}. Our method is expected to work best with the geomagnetic signal in terms of reconstruction accuracy of the radio emission profile because of the more symmetric nature of the geomagnetic fluence pattern in the shower plane compared to the total signal. A good reconstruction using the total signal would require additional corrections in order to account for their large asymmetric fluence pattern, which we plan to address in a future work. We note that the present work also do not include the effect of timing uncertainties, Galactic and instrumental noise as well as the frequency and the direction dependence of the antenna response.

\subsection{Reconstruction of arrival direction and shower core position from the total signal}
\label{subsection:core}
\begin{figure}
\includegraphics[width=1\columnwidth]{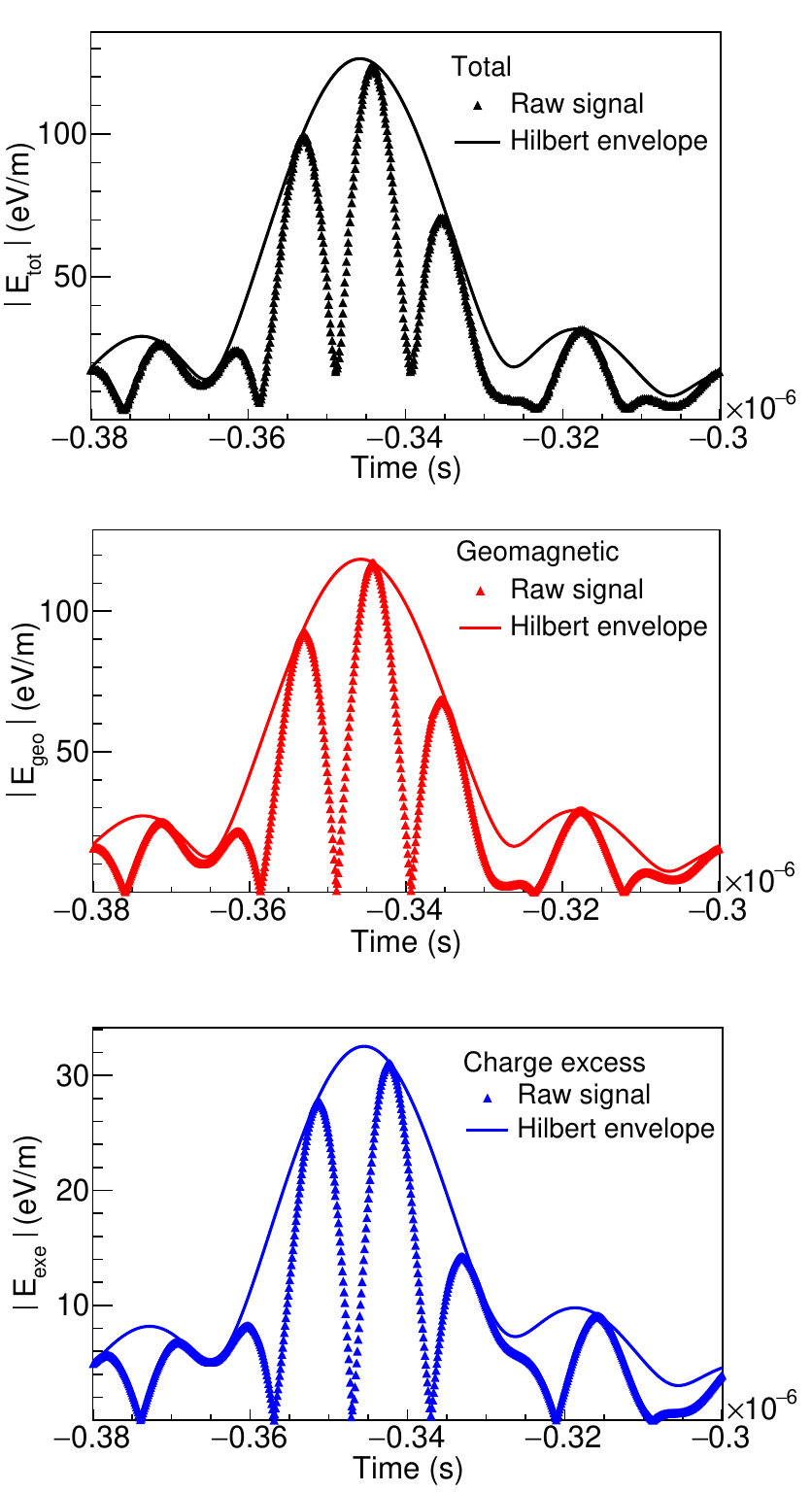}
\caption{ The time dependence of the radio signal in the frequency band of $\rm 20-80\,MHz$ for an antenna located at a distance of $\rm 28.2$~m from the shower core
  for three cases: (above $\blacktriangle$) total emission $\mid\overrightarrow{E}_{\rm tot}\mid$, (middle \textcolor{red}{$\blacktriangle$}) geomagnetic emission $\mid\overrightarrow{ E}_{\rm geo}\mid$, and (bottom \textcolor{blue}{$\blacktriangle$}) charge excess emission $\rm \mid\overrightarrow{ E}_{\rm exe}\mid$. The solid line in each panel represents the Hilbert envelope of the corresponding component. The energy of this proton EAS is $\rm 3.7\times 10^{17}\,eV$. And the arrival direction ($\rm \theta,\phi$) of the EAS is $\rm(34.3^{\circ}, 270.4^{\circ})$.}
\label{signal}
\end{figure}

For the reconstruction of the shower parameters and the subsequent finding of $X_\mathrm{max}$, we follow a procedure that is directly applicable to data from real measurements. First, we obtain the arrival time and the fluence for each antenna using the total signal obtained from simulation. The arrival time $t^\mathrm{i}$ in the $i^\mathrm {th}$ antenna is the time at which the  magnitude of the total  electric field trace $E_\mathrm{tot}^\mathrm{i}$ becomes maximum. The fluence $f_{\rm tot}^\mathrm{i}$ of the $i^\mathrm{th}$ antenna is calculated by integrating the time trace of the total electric field signal  $E_\mathrm{tot}^\mathrm{i}$ received by the antenna over a narrow time window of $\Delta t=20$~ns around the peak $t^{\rm i}$ as,
\begin{equation}
f_{\rm tot}^\mathrm{i}=\epsilon_{0}c \Delta t\sum_{k}{E_\mathrm {tot}^\mathrm{i}}^{2}(t_\mathrm{k})
\label{fluence}
\end{equation}
where $\epsilon_{0}$ is the vacuum permittivity and $c$ is the speed of light in vacuum. 
Here, the signal is sampled at $\rm 1\,ns$, prior to obtaining the arrival time  $\rm t^{i}$, and fluence $\rm f_{tot}^{i}$  in each antenna. An example of a signal trace, sampled at 1 ns time bin, along with its Hilbert envelope \cite{corstanje2023high} are shown in Figure \ref{signal}. 

In the next step, we perform an initial reconstruction of the arrival direction and the core position of the EAS using the arrival times of the total signal  received in each antenna. These initial parameters are then used to separate the geomagnetic and the charge-excess components from the total signal (see Section \ref{subsection:sep}). Finally, we use the geomagnetic signal trace to reconstruct the final values of the arrival direction and the core position using the same procedure as described in this section.

The signal arrival time $t^\mathrm{i}$ is modelled as a sum of two parts $t_\mathrm{c}^\mathrm{i}$, and $t_\mathrm{p}^\mathrm{i}$ . First part $t_\mathrm{p}^\mathrm{i}$ which is related to the propagation delay of the shower plane, and the second part $t_\mathrm{c}^\mathrm{i}$ that results from the curvature of the radio wavefront. The part $t_\mathrm{p}^\mathrm{i}$ can be described by the following equation,
\begin{equation}
c(t_{\rm p}^{\rm i}-t_{\rm 0}) =lx_{\rm i}+my_{\rm i}+nz_{\rm i}
\label{propdelay}
\end{equation}
where $(x_\mathrm{i}, y_\mathrm{i}, z_\mathrm{i})$ represent the coordinates of the antenna, $(l,m,n)$ are the direction cosines  of the EAS, $c$ is the speed of light and $t_\mathrm{0}$ represents the time at which the wavefront passes through the array center.
The direction cosines $(l,m,n)$ are related to the arrival direction $(\theta,\phi)$  of the EAS in the following manner: $l=\rm sin(\theta) cos(\phi)$, $m=\rm sin(\theta)sin(\phi)$ and $n=\rm cos(\theta)$.
The term $t_\mathrm{c}^\mathrm{i}$ corresponds to the delay of the radio wavefront with respect to the shower plane i.e the plane perpendicular to the shower axis, as shown in Fig.\ref{showerfront}. The delay is parameterized as a function of the distance from the shower core, by a fourth order polynomial function as,
\begin{equation}
ct^{\rm i}_{\rm c}=\alpha_{0}+\alpha_{ 1}r^{1}_{\rm i}+\alpha_{2}r^{2}_{\rm i}+\alpha_{3}r^{3}_{\rm i}+\alpha_{4}r^{4}_{\rm i}
\label{wavefrontdelay}
\end{equation}

Here, $r_\mathrm{i}$ is the distance of the $i^\mathrm{th}$ antenna from the shower core in the shower plane which is given by,
\begin{equation}
r_{\rm i}=\sqrt{(x_{\rm i}-X_{\rm c})^2+(y_{\rm i}-Y_{\rm c})^2-C_{\rm i}^{2}}
\end{equation}

\begin{equation}
C_{\rm i}=\sqrt{(x_{\rm i}-X_{\rm c})l+(y_{\rm i}-Y_{\rm c})m}
\end{equation}
where $(X_\mathrm{c}, Y_\mathrm{c})$ represent the coordinates of the core position at the ground. 

A fit to the arrival times $t^\mathrm{i}$ performed with this model yields the arrival direction $\rm (\theta,\phi)$, the core position $ \overrightarrow{R}_\mathrm{C}=(X_\mathrm{c},Y_\mathrm{c})$, and the parameters $\alpha_\mathrm{k}$ describing the shower front structure. The minimization is performed using the MINUIT functionality in ROOT \cite{brun1997root}. To stabilize the fit, an initial set of values for the fit parameters is determined and given as inputs to the fitting routine. The initial arrival direction is obtained by fitting the shower front with a plane function as given by Equation \ref{propdelay}. For the core position, the initial coordinates of the core position are obtained by taking the center of gravity of the fluence over the first 10 brightest antennas. The initial values of the shower curvature parameters $\alpha_\mathrm{k}$ are set to zero.


We do not explicitly include noise (both electronic and galactic noise) in our simulation. However, we exclude those antennas having an amplitude lower than a certain threshold\footnote{The threshold is set at 5\% of the maximum signal of a $10^{17}$ eV EAS having an $X_\mathrm{max}$ of $\rm 600\,g\,cm^{-2}$.} from the wavefront fitting. It is found that certain antennas with distances $\gtrsim 200$\,m from the shower core show large offsets of their arrival times $t^{i}$ from the fitted wavefront. Such outliers are removed, and the wavefront fitting is iterated to have a more accurate estimate of the arrival direction and the core position to obtain a better description of the wavefront for backtracking.

\begin{figure}
\includegraphics[width=1\columnwidth]{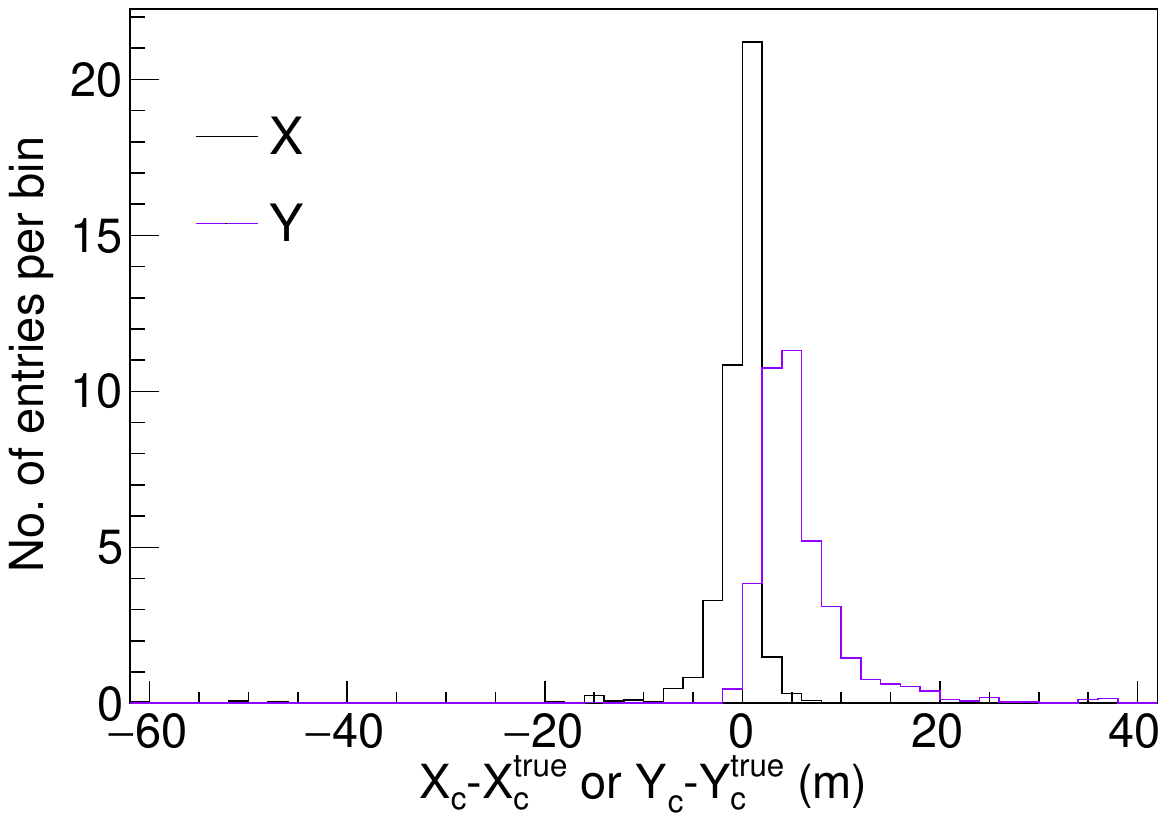}
\caption{The distribution of the difference in the reconstructed core position obtained from the radio wavefront fit  $R_{\rm c}=(X_{\rm c},Y_{\rm c})$  and the true core position $R_{\rm C}=(X_{\rm c}^{\rm true},Y_{\rm c}^{\rm true})$ for the total signal.  Here the  positive X axis is oriented towards the magnetic North, and the positive Y axis to the West.  From  the distributions, the  biases in the x-component (black) and  the y-component (violet), are obtained by fitting the peaks with a Gaussian function. The biases are estimated to be  $\rm 0.3\pm0.2\,m$  and  $\rm 4.3\pm0.5\,m$ for the  x-component and  the y-component respectively. The standard deviation of  the fitted Gaussian functions yield the resolution of the core position to be $\rm 1.1\pm0.2\,m$ and  $\rm 2.2\pm0.3\,m$ for the x-component  and  the y-component respectively. The distributions are for proton EAS in the zenith angle range of $\rm 0^{\circ}-10^{\circ}$ and energy range of $\rm 10^{17}-10^{17.5}\,eV$. The true core positions are located at the center of the array.} 
\label{cogtimecore}
\end{figure}

The core position is estimated from the radio wavefront, unlike the commonly followed procedure of fitting the fluence pattern at the ground. The distribution of the difference in the reconstructed and true core positions for $\rm 10^{17}-10^{17.5}\,eV$ proton EAS in the zenith angle bin of $0^{\circ}-10^{\circ}$ are shown in Figure \,\ref{cogtimecore}. The bias in the core position $R_{\mathrm c}$ obtained from a Gaussian fit around the peak of the distribution are found to be $\rm 0.3\pm0.2\,m$  and  $\rm 4.3\pm0.5\,m$ along the $X$-axis (black) and $Y$-axis  (violet) respectively. The resolution of the core position $R_{\mathrm c}$ as obtained from the Gaussian fit is ($\rm 1.1\pm0.2\,m$, $\rm 2.2\pm0.3\,m$).

\subsection{Separation of geomagnetic and charge-excess components}
\label{subsection:sep}
The arrival direction ($\theta$,$\phi$), and the core position $\overrightarrow{R}_C$   obtained from the subsequent fit are then used to separate the  total signal $E_{\rm tot}$ into the geomagnetic component $\rm \overrightarrow{E}_\mathrm{geo}$ and the charge excess component $\rm \overrightarrow{E}_\mathrm{exe}$. 
To a good accuracy, the geomagnetic and the charge excess components of the electric field $\rm \overrightarrow{ E}_\mathrm{geo}$ and $\rm \overrightarrow{E}_\mathrm{exe}$, at a  position $\overrightarrow{r}$ from the shower core and time t can be written as the sum of the electric field components along the $\overrightarrow{V}\times\overrightarrow{B}$, and the $\overrightarrow{V}\times(\overrightarrow{V}\times\overrightarrow{B})$ directions as 
\begin{equation}
E_\mathrm{geo}(\overrightarrow{r},t)=E_{\overrightarrow{\rm V}\times\overrightarrow{\rm B}}(\overrightarrow{r},t)-\frac{\cos \delta}{\sin {\delta}}E_{\overrightarrow{\rm V}\times(\overrightarrow{\rm V}\times\overrightarrow{\rm B})}(\overrightarrow{r},t)
\label{sep1}
\end{equation}
\begin{equation}
E_\mathrm{exe}(\overrightarrow {r},t)=\frac{1}{\sin \delta}E_{\overrightarrow{\rm V}\times(\overrightarrow{\rm V}\times\overrightarrow{\rm B})}(\overrightarrow{r},t)
\label{sep2}
\end{equation}
where $\delta$ is the polar angle between the observer position and the $\vec{V}\times \vec{B}$ axis. Equations \ref{sep1} and \ref{sep2} are valid under the assumption that both the emission components arrive at the antenna simultaneously without any phase difference \cite{glaser2016simulation}. We neglect the small phase difference between the two components as indicated by the  observations of circular polarization in radio emission with LOFAR \cite{scholten2016measurement}. The average phase difference is found to be $\rm \lesssim 1\,ns$ within and around the Cherenkov ring, which is negligible compared to the total pulse width of tens of nanoseconds \cite{schluter2020refractive}. Moreover, in the present study, we do not consider the non-negligible phase difference that may be present at regions far away from the Cherenkov ring. A proper treatment of this will be included in a future study that will focus on an improvement over this proof of principle study. It can be noticed that Equations \ref{sep1} and \ref{sep2} diverges due to the term $\rm \frac{1}{sin(\delta)}$  when the antenna position is close to the $\rm \overrightarrow V\times \overrightarrow B$ axis. In order to handle this issue, only antennas located outside $\rm 15^{\circ}$ from the $\rm \overrightarrow V\times \overrightarrow B$ axis are considered for the reconstruction purposes. 

\begin{figure}
\includegraphics[width=1\columnwidth]{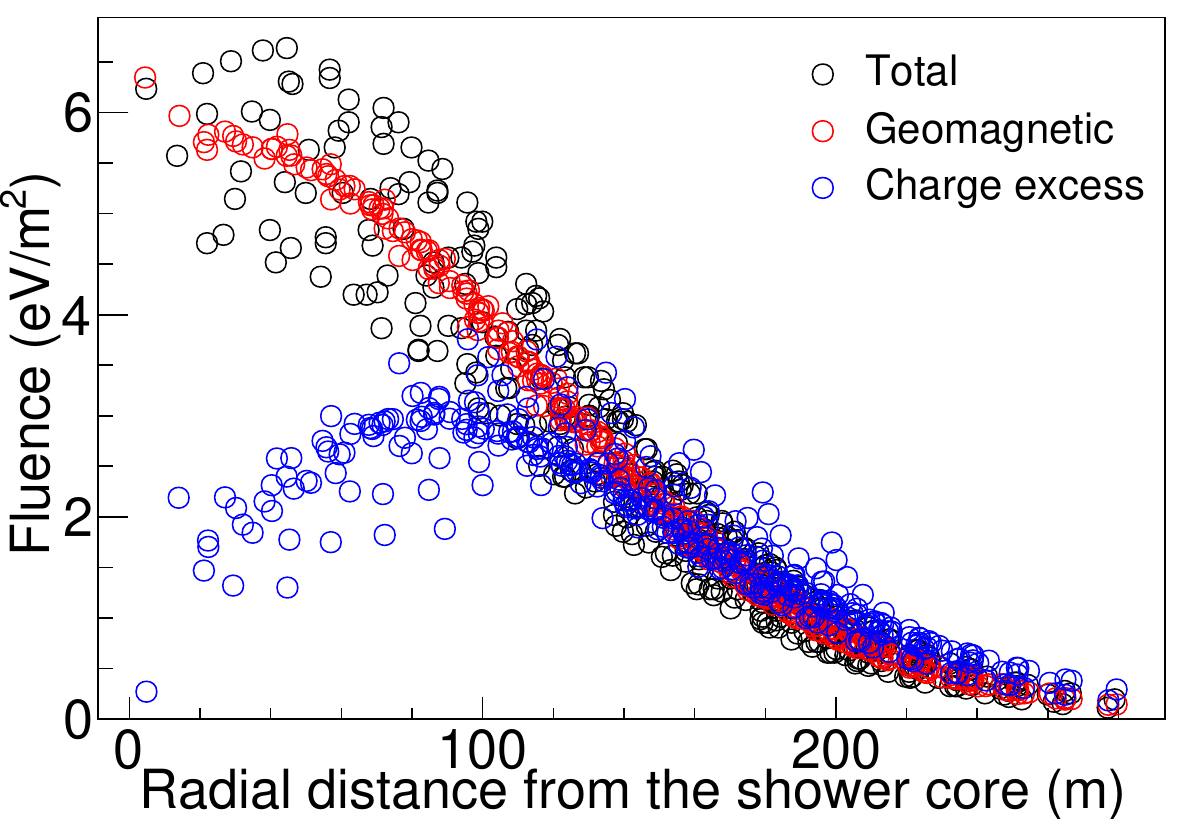}
\caption{Comparing the lateral distribution (variation in the fluences $f$ as a function of the distance from the EAS core in the shower plane, $r$) of the geomagnetic signal (red), with that of the total signal (black) for the frequency band of $\rm 20-80\,MHz$. A significant reduction in the scatter is observed on obtaining  the geomagnetic component. The scatter (root mean square) of the fluence for the geomagnetic signal $f_{\rm geo}$ at $\rm 20-30\,m$ is 1.1\% which is a significant reduction when compared to its value of 13.7\% for the  total signal $f_{\rm tot}$. In the region of $\rm 200-210\,m$, the scatter of the fluence for the total and geomagnetic signals are 21.5\% and 12.4\%, respectively. The energy of this proton EAS is $\rm 8.3\times 10^{17}\,eV$. The EAS arrival direction ($\rm \theta,\phi$) is $\rm (30.3^{\circ},196.5^{\circ})$. The fluence of the charge excess signal $f_{\rm exe}$ (blue) has been scaled by a factor 35. The scatter for the charge excess emission is $\rm 17\%$ at $\rm 20-30\,m$.}
\label{ldf}
\end{figure}

The origin of the $(\overrightarrow{V} \times \overrightarrow{B}, \vec{V}\times (\vec{V}\times \vec{B}))$  coordinate system is positioned to match the reconstructed core position $\overrightarrow{\rm R}_{\rm C}$ projected onto the shower plane. Fig \ref{signal} shows an example of the total, geomagnetic and charge excess signals received by an antenna located at a distance of $28.2$~m from the array center for a proton shower of $\rm 5.4\times10^{17} eV$. The amplitude of the geomagnetic signal is $\rm 114.1\,\upmu V/m$ for this antenna, which is $\sim$ 4 times that of the charge excess signal.

The fluence of the geomagnetic and the charge-excess components of each antenna can be obtained separately from the time trace of the respective signals using Equation \ref{fluence}. Figure~\ref{ldf} shows the one-dimensional lateral distribution of the geomagnetic fluence $f_{\rm geo}$ constructed in the shower plane together with the distribution of the total signal for a proton shower of energy $8.2\times 10^{17}$~eV and zenith angle $\theta=30.3^\circ$. The fluence of the  charge-excess component $f_{\rm exe}$ is $\sim 35$ times smaller than the geomagnetic fluence for the given shower, and its scaled lateral distribution is shown in the same figure. Compared to the geomagnetic component, both the charge-excess component and the total signal are found to show a much large scatter in the lateral distribution.  At a distance of $\rm 20-30\,m$ from $\rm \overrightarrow{\rm R}_{C}$ in the shower plane, the spread (root mean square) of the fluence for the total emission is $\sim~13.7\%$, which increases to $\sim~21.5\%$ at $200-210$~m. The scatter for the charge-excess component is even larger reaching about $17\%$ at $20-30$~m. On the other hand, the spread for the geomagnetic signal is significantly less which is only about $1.1\%$ at the same distance of $20-30$~m from the shower core.

\begin{figure}
\includegraphics[width=1\columnwidth]{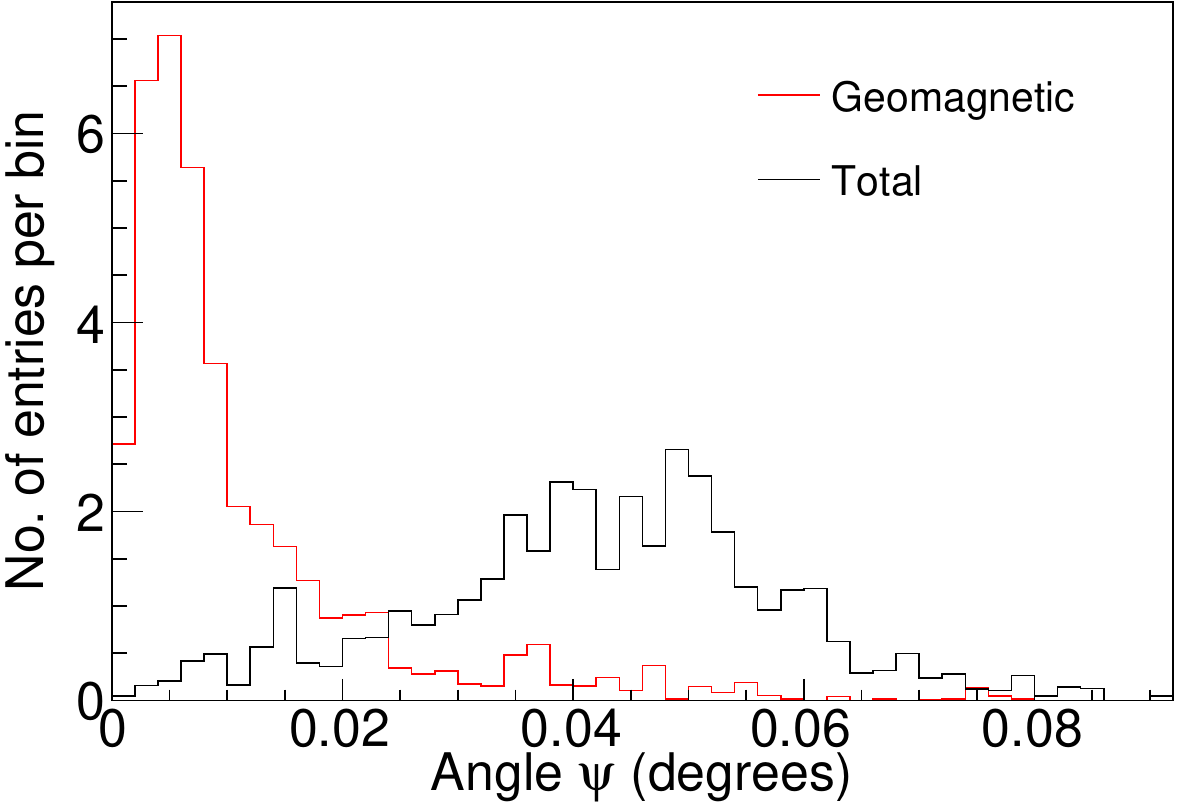}
\caption{The distribution of the separation angle between the true and reconstructed arrival directions for the geomagnetic emission (red) and the total signal (black) for proton EAS in the zenith angle range of $0^{\circ}-10^{\circ}$ and energy range of $\rm 10^{17}-10^{17.5}\,eV$ (same showers as used in Figure\,\ref{cogtimecore}). The angular resolution, defined as the $68\%$ containment region, are obtained as $\rm 0.01^{\circ}$ and $\rm 0.04^{\circ}$ for the geomagnetic and the total signals respectively.}
\label{anglecomp}
\end{figure}

\begin{figure}
\includegraphics[width=1\columnwidth]{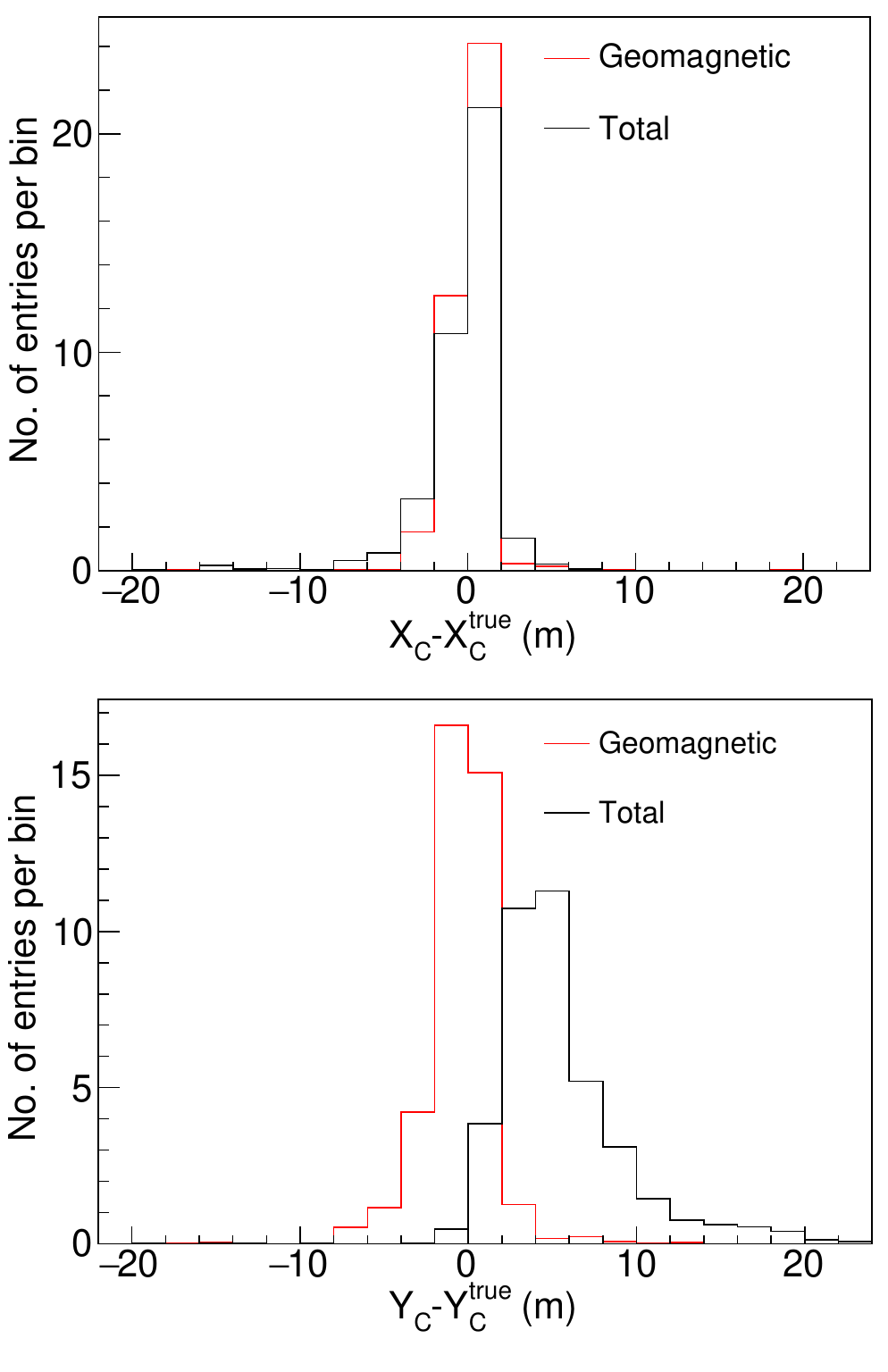}
\caption{The distribution of the difference in the reconstructed core position  $R_{\rm C}=(X_{\rm C},Y_{\rm C})$, and the true core position $R_{\rm C}=(X_{\rm C}^{\rm true},Y_{\rm C}^{\rm true})$, for two cases: 1) geomagnetic signal (red), 2) total signal (black), in the zenith angle bin of $\rm 0^{\circ}-10^{\circ}$. From  the distributions, the resolution of  the x-component (above) and  the y-component (below), are obtained by fitting them with a Gaussian function. Here the  positive X axis is oriented towards the magnetic North, and the positive Y axis to the west. The resolution of the core position $R_{\rm C}=(X_{\rm C},Y_{\rm C})$ is $\rm (-0.9\pm0.2\,m,1.5\pm0.2\,m)$, and $\rm(1.1\pm0.2\,m,\rm 2.2\pm0.3\,m)$ for geomagnetic and the total signal respectively. The fitted Gaussian function peaks at ($\rm0.4\pm0.2\,m,4.5\pm0.5\,m$) for the total signal, whereas for geomagnetic signal the peak occurs at ($\rm0.3\pm0.2\,m, -0.3\pm0.1\,m$).  The distributions are that of proton EAS, in the zenith angle bin of $0^{\circ}-10^{\circ}$, in the energy range of $\rm 10^{17}-10^{17.5}\,eV$. The distribution here is of the same ensemble of EAS shown in Figure \ref{cogtimecore}.} 
\label{corecomp}
\end{figure}

\subsection{{Reconstruction of EAS radio emission profile using geomagnetic signal}}
\label{reco}
Having extracted the geomagnetic signal $\rm \overrightarrow{\rm E}_{geo}$, the arrival direction and the core position are reconstructed using purely the geomagnetic wavefront following a similar procedure described in the Section\,\ref{subsection:core}. A comparison of the distribution of the angle between the true and the reconstructed arrival direction $\psi$ between the geomagnetic and the total signal is shown in the Figure~\ref{anglecomp} for proton EAS of $\rm 10^{17}-10^{17.5}\,eV$ in the zenith angle range of $\rm 0^{\circ}-10^{\circ}$. Assuming a Gaussian distribution and defining the angular resolution as the 68\% containment region, the angular resolution
is obtained  as $\rm 0.01^{\circ}$, and $\rm 0.05^{\circ}$ for geomagnetic and the total signal respectively. 

\begin{figure*}
\includegraphics[width=1\textwidth]{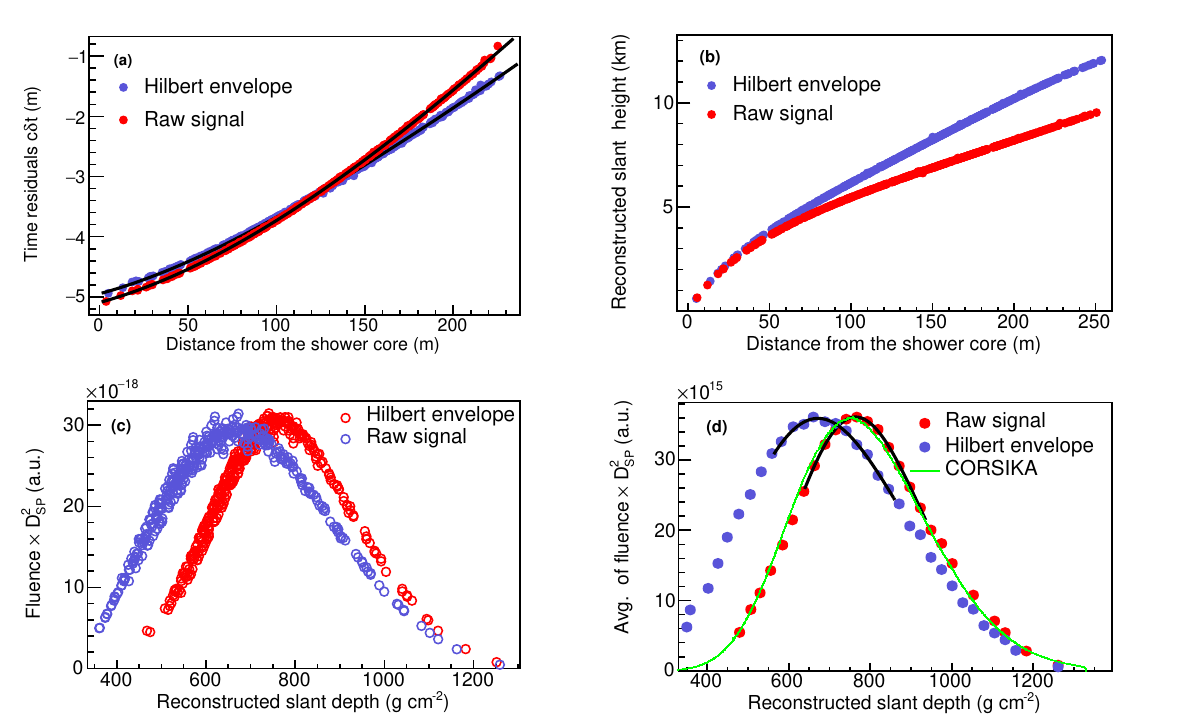}
\caption{(a) The geomagnetic wavefront: time residuals $c\delta t$ as a function of the distance $r$ from the shower core in the shower plane for the raw signal (red) and the Hilbert envelope (blue) cases. The black lines represent the fourth order polynomial fits, represented by equation\,\ref{wavefrontdelay}, on the respective wavefront. (b) The variation of the reconstructed slant height  $H_{\rm SP}=\sqrt{D_{\rm SP}^{2}-{r^{2}}}$ as a function of the distance from the shower core $r$ for the raw signal (red) and Hilbert envelope (blue). (c) The plot of ($ f_{\rm geo}\times D^2_{\rm SP}$) as a function of the slant depth for the raw signal (red) and the Hilbert envelope (blue). (d) The red dots represent the reconstructed longitudinal profile (the average of the  $f_{\rm geo}\times D^2_{\rm SP}$,  over $\rm 26\,g\,cm^{-2}$, as a function of the slant depth) for the raw signal (red) and Hilbert envelope (blue). The reconstructed longitudinal profiles are scaled so that its Y-value at the peak matches that of the true longitudinal profile $N_\mathrm{e}^{2}$ (green line), obtained from CORSIKA. The black line is the fitted Gaisser Hillas function on the reconstructed profile, the peak of which gives the $R_\mathrm{max}$ to be $\rm 764.1\,\pm\,1.7\,g\,cm^{-2}$ for the raw signal. However for the Hilbert envelope case, the peak is estimated to be $\rm 671.7\,\pm 2.41\,g\,cm^{-2}$. The  $X_\mathrm{max}$ of this EAS  is $\rm 753.5\,g\,cm^{-2}$. The arrival direction $\rm(\theta,\phi)$, and the energy of this proton EAS are $\rm (37.8^\circ,154.7^\circ)$, and $\rm 2.5\times10^{17}\,eV$.}
\label{longi}
\end{figure*}

For the same showers used in Figure~\ref{anglecomp}, we show in Figure~\ref{corecomp}, a comparison of the distribution of the reconstructed core position with respect to the true position ($X_\mathrm{C}-X_\mathrm{C}^\mathrm{true}$ and $Y_\mathrm{C}-Y_\mathrm{C}^\mathrm{true}$) obtained from the geomagnetic signal with that obtained using the total signal. Although there is no noticeable difference observed in the $X_\mathrm{C}-X_\mathrm{C}^\mathrm{true}$ distributions, there is a significant reduction in the bias of the $Y_\mathrm{C}-Y_\mathrm{C}^\mathrm{true}$ distribution for the geomagnetic signal. A Gaussian fit to the distributions yields the core resolution $(\sigma_\mathrm{X},\sigma_\mathrm{Y})$ to be $\rm (-0.9\pm0.2\,m,1.5\pm0.2\,m)$ and $\rm(1.1\pm0.2\,m,\rm 2.2\pm0.3\,m)$ for the geomagnetic and the charge-excess signal respectively. The bias obtained from the Gaussian peaks of the $Y_\mathrm{C}-Y_\mathrm{C}^\mathrm{true}$ fits are found to be $\rm -0.3\pm0.1\,\,m$ for the geomagnetic emission and $\rm 4.5\pm0.5\,m$ for the total signal.


Using the reconstructed core position and the arrival direction, the wavefront $c\delta t=ct-ct_{\rm p}$  of the geomagnetic emission is constructed as a function of $ r$, as shown in Figure~\ref{longi}a for a proton EAS of $\rm 2.5\times10^{8}\,eV$. The red dots represent the wavefront obtained using the raw signal and the blue dots represent the Hilbert envelope case. A fourth order polynomial fit, as represented by Equation~\ref{wavefrontdelay}, is performed on the wavefront. The derivative of the fitted polynomial at a given point  gives the slope of the wavefront.
In the next step,  the effective source point for each antenna along the shower axis is obtained. This is performed by constructing straight lines emanating from the antenna position and passing through the wavefront, then selecting the line that is perpendicular to the wavefront. To save computational time, the check for perpendicularity, is performed only for those lines that intersect the wavefront within a radial distance  of $\rm 2\,m$ on the either side of the antenna in the shower plane. Such a cut was imposed based on the observation that the perpendicular line, always intersects the wavefront within this distance from the antenna. The effective source point is given by the point of intersection of the constructed perpendicular line with the shower axis. Figure~\ref{longi}b shows the reconstructed slant height  of the effective source points as a function of the  distance from the shower core in the shower plane for the raw signal (red) and Hilbert envelope (blue) cases.  

The reconstructed height associated with each antenna is converted into the corresponding slant depth $X$, based on the GDAS model parameter originally used to produce the simulated EAS \cite{mitra2020reconstructing}.  Assuming that the radio emission from EAS is strongly beamed in the forward direction within a narrow cone, the geomagnetic emission at a source point ($S^\mathrm{i}_\mathrm{geo}$) is determined by scaling the geomagnetic fluence ($f^\mathrm{i}_\mathrm{geo}$) recorded at an antenna with the  square of the distance of the source point from the antenna position at the ground (${D_\mathrm{SP}^\mathrm{i}}$).  $S^\mathrm{i}_\mathrm{geo}$ is also expected to be proportional to the square of the total number of electrons and positrons (${N_\mathrm{e}}^{2}$) contained in the EAS \cite{huege2003radio}. Figure~\ref{longi}c shows the values of the calculated $S^\mathrm{i}_\mathrm{geo}$ as a function of the slant depth $X$ for the same event shown in Figures~\ref{longi}a and \ref{longi}b. The radio emission profile of the shower is then finally obtained from Figure~\ref{longi}c by averaging $S^\mathrm{i}_\mathrm{geo}$ over $X$ bins. Such a reconstructed radio profile is shown in Figure~\ref{longi}d for an $X$ bin size of $\rm 26\,g\,cm^{-2}$ for both the raw and the Hilbert envelope cases. To compare with the longitudinal profile of the EAS, the distribution of ${N_{e}}^{2}$ obtained from CORSIKA simulation for the same event is also shown in Figure~\ref{longi}d (green line).

The reconstructed radio emission profile based on the raw signal (red dots) is found to agree quite well with the shower longitudinal profile. Fitting the reconstructed profile with the Gaisser-Hillas function \cite{gaisser1977reliability} gives the position of the peak of the radio profile at $R_\mathrm{max}=\rm 764.1\,\pm1.7\,g\,cm^{-2}$. This value differs only by $\rm 10.6\,g\,cm^{-2}$ from the $X_\mathrm{max}$ value of the shower which is at $\rm 753.5\,g\,cm^{-2}$. For the same EAS, the radio emission profiles constructed using the total and the charge-excess signals, following a similar procedure, are shown in Appendix\,\ref{app1} for the raw signal case. 

On the other hand, the radio emission profile obtained using the Hilbert envelope (blue dots in Figure~\ref{longi}d) shows a large systematic offset with respect to the longitudinal profile of the shower. The reconstructed $R_\mathrm{max}$ is found to be shifted by about $\rm 80\,g\,cm^{-2}$ from the $X_\mathrm{max}$ value in this case. This is not due to bad reconstruction of the arrival direction or the shower core position. For the shower shown in Figure~\ref{longi}, the reconstructed arrival direction from the Hilbert envelope and the raw signal cases differs by only about $\rm 4\times 10^{-4}\,deg$, and the reconstructed core position by only $\sim\rm 2\,m$. The observed offset is found to be related to the less concave nature of the shower wavefront constructed using the Hilbert envelope (Figure~\ref{longi}a) which shows a deviation from the raw-signal wavefront that increases progressively with the distance from the shower core.
At a distance of $\rm 10\,m$ from $R_{\rm C}$, in the shower plane, the time residual $c\delta t$  of the Hilbert envelope differs by $\sim \rm 0.2\,m$, whereas at a distance of $\rm 200\,m$, it differs by $\sim \rm -0.3\,m$ as shown in Figure~\ref{longi}a. As a result the reconstructed height also differ significantly from that obtained from the raw signal. The difference in the reconstructed height is $\sim \rm 0.5\,km$ at $\rm 80\,m$ which increases to $\sim \rm 2\,km$ at $\rm 200\,m$ as shown in Figure~\ref{longi}b.
 
We note that the large systematic offset observed in $R_\mathrm{max}$ for the Hilbert envelope case is found to be present for all the showers that we have analysed for the present work. This discrepancy needs further investigation, which will be the focus of a follow-up work in future. Hereafter, for the rest of the results presented in this paper, we will only focus on the results obtained using the raw signal.


In Appendix\,\ref{app2}, we have shown several examples of radio emission profiles of proton and iron showers obtained using geomagnetic signal for the zenith angle range of $0^{\circ}-40^{\circ}$ and $X_\mathrm{max}$ range of $\sim\,\rm 500-950\,g\,cm^{-2}$. For some EAS, the wavefront cannot be well reconstructed due to the poor reconstruction of the arrival direction and the shower core position. Those events are removed from the analysis. They are identified based on the non-convergent fit of the wavefront or from the minimum number of antennas required to successfully obtain the reconstruction heights which is set to 5 antennas in the present study .  Using these two criteria, a reconstruction efficiency of $\sim 97\%$ is obtained, in the zenith angle range of $0^{\circ}-40^{\circ}$.

\begin{figure}
\includegraphics[width=0.53\textwidth]{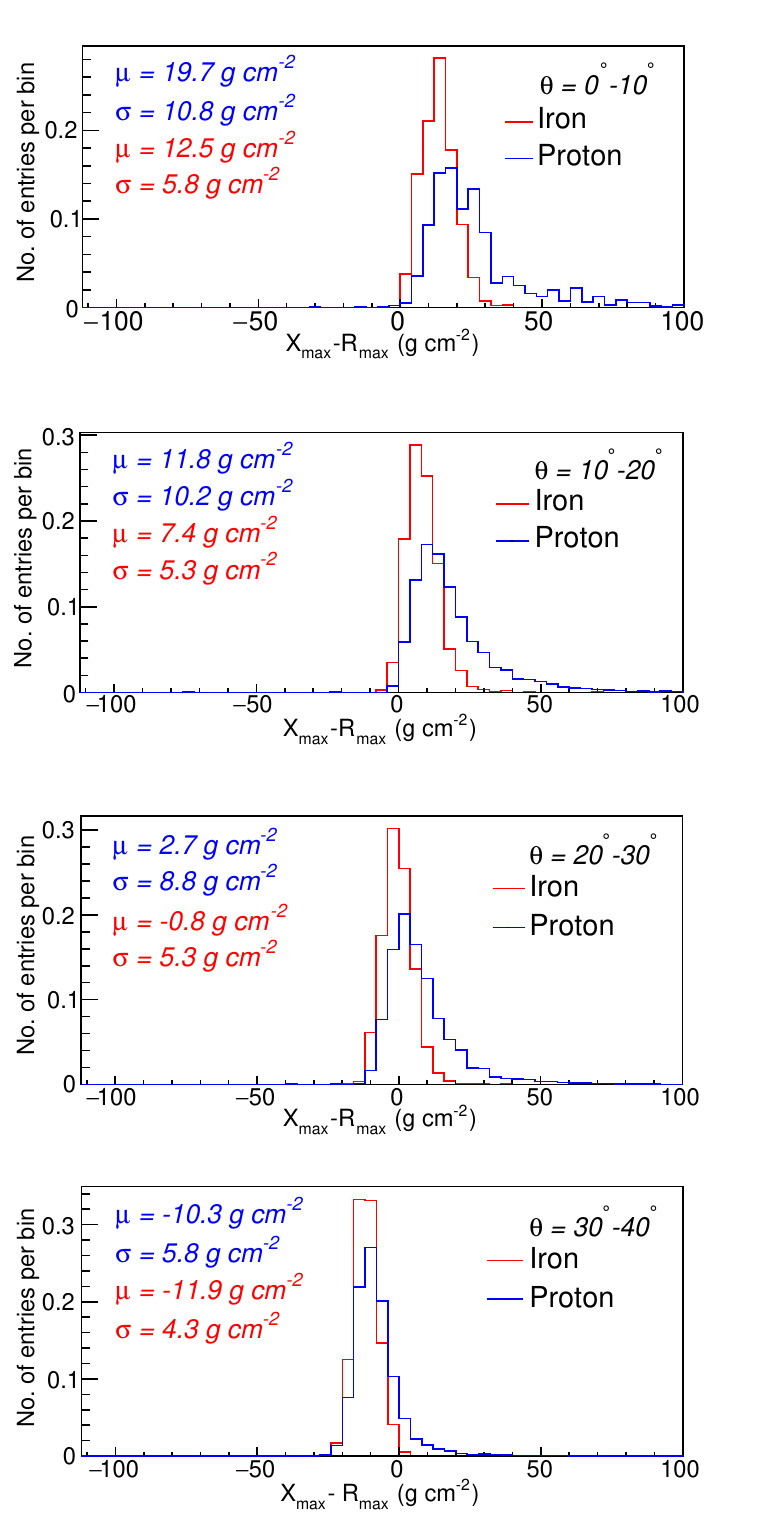}
\caption{$X_\mathrm{max}-R_\mathrm{max}$ distribution (normalized to one) for the proton (blue) and iron EAS (red) in the energy range of $\rm 10^{17}-10^{17.5}\,eV$ for four different zenith bins: $\rm 0^{\circ}-10^{\circ}$, $\rm 10^{\circ}-20^{\circ}$, $\rm 20^{\circ}-30^{\circ}$ and $\rm 30^{\circ}-40^{\circ}$. The distributions are obtained without including noise and electronic response in the analysis. The bias $\mu$ are determined by fitting a Gaussian function around the peak of the distribution, and the spread $\sigma$ represent the $68\%$ containment region from the peak. Note that the $\rm 68\%$ containment region is determined by using a finer bin width of $\rm 0.5\,g\,cm^{-2}$.
}
\label{figprotoniron}
\end{figure}

\section{{Comparison of $R_\mathrm{max}$ from the reconstructed radio emission profile with shower $X_\mathrm{max}$}}
In order to check the potential of the method for $X_\mathrm{max}$ reconstruction, we apply the method to a sample of proton and iron showers in the energy range of $\rm 10^{17}-10^{18}\,eV$ and zenith angle range of $0^{\circ}-40^{\circ}$, and compare the $R_\mathrm{max}$ values obtained from the reconstructed radio emission profile with the $X_\mathrm{max}$ values from CORSIKA. Figure\,\ref{figprotoniron} shows the  distribution of the difference between the $X_\mathrm{max}$ and $R_\mathrm{max}$ values for proton and iron showers obtained for four separate zenith angle bins of $0^{\circ}-10^{\circ}$, $10^{\circ}-20^{\circ}$, $20^{\circ}-30^{\circ}$ and $30^{\circ}-40^{\circ}$ in the energy range of $\rm 10^{17}\,eV-10^{17.5}\,eV$. The $\sigma$ value (spread) of the distribution which is taken as the $68\%$ containment region from the peak and the bias value $\mu$ which is obtained from a Gaussian fit around the peak are listed in Table~\ref{resbias}. The table also provides the $\sigma$ and $\mu$ values for the $\rm 10^{17.5}-10^{18}\,eV$ energy range whose corresponding plots are shown in Appendix \ref{appendixC} (Figure \ref{xmaxdiff2}).

In the $\rm 10^{17}-10^{17.5}\,eV$ range, the $\sigma$ value for protons is $\rm\sim~11\,g\,cm^{-2}$ at $\rm 0^{\circ}-10^{\circ}$ zenith angle, and the value gets smaller at higher zenith angles, reaching $\rm\sim~6\,g\,cm^{-2}$ at $\rm 30^{\circ}-40^{\circ}$. This decrease in $\sigma$ is due to the increase in the distance to the shower maximum as the zenith angle increases. Showers with $X_\mathrm{max}$ located closer to the ground are generally poorly reconstructed as shown in Appendix\,\ref{app2} (Figures \ref{longiprof_20} and \ref{longiprof_20_iron}). For the iron showers, the $\sigma$ values are relatively smaller by a factor of $\sim\,1.3-1.9$ depending on the zenith angle, and they show a similar trend with zenith angle as the proton showers. At  $\rm 10^{17.5}-10^{18}\,eV$, the $\sigma$ values for both the proton and iron showers are in general found to be slightly larger than at $\rm 10^{17}-10^{17.5}\,eV$ by a factor of $\rm \sim 1.1-1.3$. This is expected as higher energy showers penetrate deeper in the atmosphere leading to $X_\mathrm{max}$ values closer to the ground. However, in the presence of noise, the $\sigma$ values at higher energies mentioned above can be reduced due to the increase in the signal-to-noise ratio of the radio signals with energy.

\begin{figure}
\includegraphics[width=1\columnwidth]{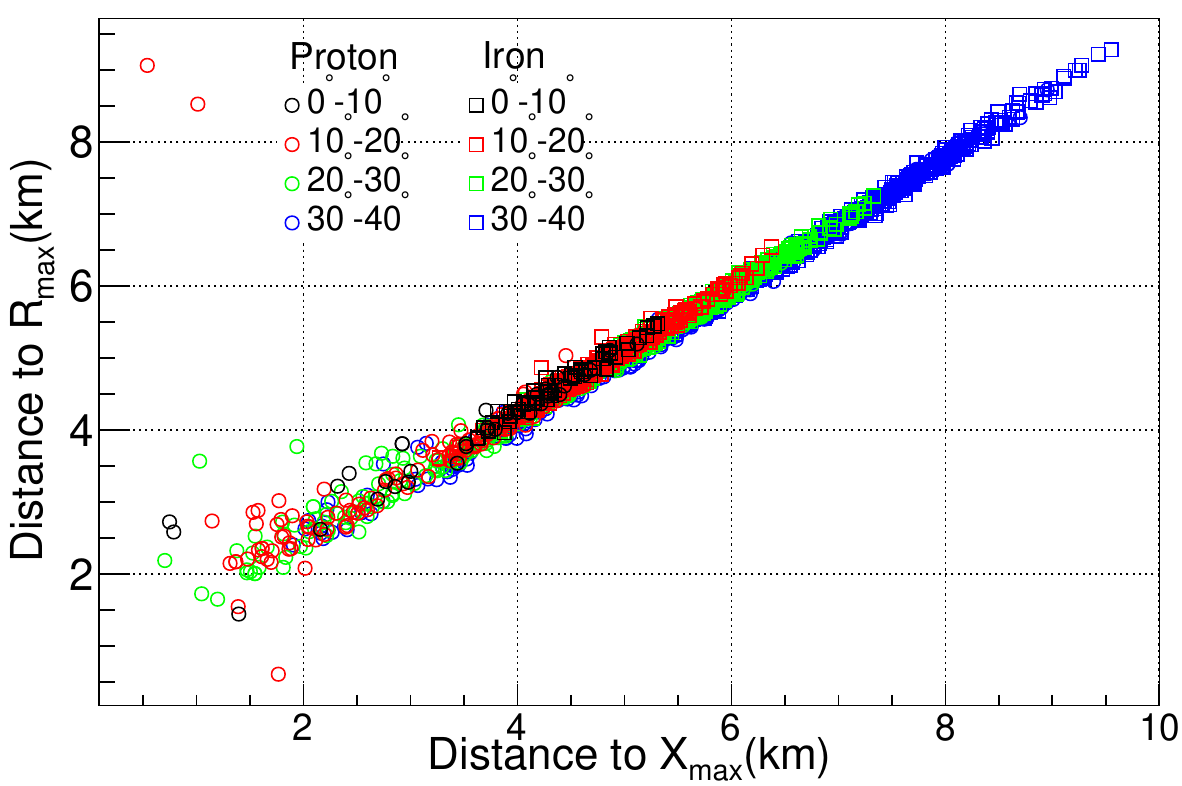}
\caption{Plot of the distance to the reconstructed $R_\mathrm{max}$ versus $d_\mathrm{max}$ (distance to $X_\mathrm{max}$) for different zenith angle bins: $\rm 0^{\circ}-10^{\circ}$ (black), $\rm 10^{\circ}-20^{\circ}$ (red), $\rm 20^{\circ}-30^{\circ}$ (green), $30^{\circ}-40^{\circ}$ (blue). The markers $\rm \circ$ and $\rm\square$ represent proton and iron EAS respectively. For similar values of $d_\mathrm{max}$, the plots show similar  biases for the different zenith angle bins. }
\label{scatter}
\end{figure}

The bias $\mu$ is also observed to vary with the zenith angle, the best value being observed in the zenith bin of $\rm 20^{\circ}-30^{\circ}$. For proton EAS at $\rm 10^{17}\,eV-10^{17.5}\,eV$ energies, the bias is $\rm 19.1\,g\,cm^{-2}$ at $\rm 0^{\circ}-10^{\circ}$, reaching a minimum value of $\rm 2.7\,g\,cm^{-2}$ at $\rm 20^{\circ}-30^{\circ}$ which is also the zenith region where majority of the showers are observed in actual measurements \cite{thoudam2014lora} and increases to $\rm -10.3\,g\,cm^{-2}$ at $\rm 30^{\circ}-40^{\circ}$. We have found that the change in the bias with zenith angle is related to the change in the average distance to $X_\mathrm{max}$. In general, inclined showers have $X_\mathrm{max}$ values higher up in the atmosphere compared to the less inclined ones. For showers with similar distances ($d_\mathrm{max}$) to $X_\mathrm{max}$, we find the bias to be comparable between different zenith angle bins as shown in Figure\,\ref{scatter}, indicating that the change in the bias is due to the change in $d_\mathrm{max}$. At large $d_\mathrm{max}$ (small $X_\mathrm{max}$) which corresponds to higher zenith angles, the distance to the reconstructed $R_\mathrm{max}$ is smaller than the $d_\mathrm{max}$ value as can be seen in Figure~\ref{scatter}. An opposite trend is observed for showers at lower zenith angles (smaller $d_\mathrm{max}$ or larger $X_\mathrm{max}$). This is reflected as a shift in the plots from right to left in Figure \ref{figprotoniron} as zenith angle increases.


\begin{table*}
\caption{Values of the spread $\sigma$ and bias $\mu$ (in $\rm g\,cm^{-2}$) of the $X_\mathrm{max}-R_\mathrm{max}$ distribution for proton and iron EAS considered in this work. The values are obtained without including noise and electronic response in the analysis.}
\label{resbias}
\centering
\begin{tabular}{*{9}{c}}
\hline
zenith angle & \multicolumn{4}{|c|}{$\rm E\,=\,10^{17}\,eV-10^{17.5}\,eV$} & \multicolumn{4}{c}{$\rm E\,=\,10^{17.5}\,eV-10^{18}\,eV$}\\
\cline{2-9}

$\rm \uptheta$& \multicolumn{2}{|c}{proton} & \multicolumn{2}{|c|}{iron} & 
\multicolumn{2}{|c}{proton} & \multicolumn{2}{|c}{iron} \\
\cline{2-9}

& \multicolumn{1}{|c}{$\sigma$} & \multicolumn{1}{|c|}{$\mu$} &
\multicolumn{1}{|c}{$\sigma$} & \multicolumn{1}{|c|}{$\mu$} &
\multicolumn{1}{|c}{$\sigma$} & \multicolumn{1}{|c|}{$\mu$} &
\multicolumn{1}{|c}{$\sigma$} & \multicolumn{1}{|c}{$\mu$}\\ 
\hline
$\rm 0^{\circ}-10^{\circ}$ & 10.8 & 19.7& 5.8& 12.5& 13.8& 27.4& 5.3& 17.6\\
$\rm 10^{\circ}-20^{\circ}$ &10.2&11.8&5.3&7.4&12.3&19.3&6.3&12.4\\
$\rm 20^{\circ}-30^{\circ}$ &8.8&2.7&5.3&-0.8&9.8&6.1&5.8&0.6\\
$\rm 30^{\circ}-40^{\circ}$ &5.8&-10.3&4.3&-11.9&7.3&-7.4&4.8&-10.2\\
\hline
\end{tabular}
\end{table*}



\section{Discussion}
The bias in the $(X_\mathrm{max}-R_\mathrm{max})$ distribution can be attributed to the uncertainties associated with the determination of the position of the effective source points and the emitted signals which, in turn, depend on the quality of the reconstructed shower parameters such as the core position and the arrival direction as well as on accuracy of the distance of the effective source point from the antenna position among others as described below. A bias of $\sim \rm 5\,g\,cm^{-2}$ arises as a result of the inherent bias in the core position obtained from the total radio wavefront which is then used to extract the geomagnetic signal from the total emission. An additional round of extraction of the geomagnetic signal using the core position (and the arrival direction) obtained from the geomagnetic wavefront can reduce this bias.


Additional bias may arise from the effect of the refractive index of the air on the signal trajectory. In the present  study, we have not taken into account the bending of the ray due to the varying refractive index of the air while backtracking the signal \cite{schluter2020refractive}. This can lead to an inaccurate determination of the position of the effective source point along the shower axis.

Another artifact that needs further investigation is the reconstruction of EAS having $X_\mathrm{max}$ close to the ground, i.e., $d_\mathrm{max}\,\lesssim\,2$\,km. For such showers, the reconstructed $R_\mathrm{max}$ are found to show a large systematic shift from the $X_\mathrm{max}$ value. Some examples of such EAS can be seen in Appendix\,\ref{app2} (fourth and fifth panels on the left column in Figure \ref{longiprof_20}).  For the zenith angle bin of $0^{\circ}-10^{\circ}$ and energy range of $\rm 10^{17}\,eV-10^{17.5}\,eV$, the reconstructed $R_\mathrm{max}$ of such showers show an offset of $\mu\sim \rm 90\,g\,cm^{-2}$. We have checked that this large offset is not due to  a poor reconstruction in the arrival direction or the core position of the shower. The offset could be related to the far-field approximation for the polarization of the geomagnetic and the charge excess components considered in the present study. It could also be an effect resulting from the dependence of the electric field strength on the direction of emission of the signal with respect to the shower axis. The far-field may not represent a good approximation for showers with a large amount of radiation produced close to the ground.

For the implementation of this technique, the frequency range of $\rm 20-80\, MHz $ is preferred. The Cherenkov like effects in radio emission  lead to simultaneous arrival of signals emitted at different times producing a ring-like structure in the footprint pattern at frequencies $\rm \gtrsim\,100\, MHz $. This reduces the accuracy of the method at frequencies above $\rm \sim\,100\, MHz $. For instance, for the frequency window of $\rm 100-300\,MHz$, the EAS presented in Figure \ref{longi} is poorly reconstructed, yielding an $R_\mathrm{max}$ value that is far offset from the $X_\mathrm{max}$ value by $\sim \rm 50\,gcm^{-2}$. However, as demonstrated already, $R_\mathrm{max}$ for the same event for $\rm 20-80\, MHz$ differs only by $\sim \rm 10\,g\,cm^{-2}$ from $X_\mathrm{max}$.



The angular resolution of $\rm\sim\,0.01^{\circ}$ and the $X_\mathrm{max}$ resolution ($\sigma$ of the $X_\mathrm{max}-R_\mathrm{max}$ distribution) of $\rm \lesssim 14\,g\,cm^{-2}$ are based on an idealized set-up using a simple geometry of the array, and without including noise and antenna response. We have also explored the influence of the antenna spacing on the $X_\mathrm{max}$ resolution. For the antenna spacing of $\rm 20\,m$ adopted in this paper, a resolution of $\rm 3.1\,g\,cm^{-2}$ for $\rm 2.5\times 10^{17}\,eV$ proton EAS has been obtained at a zenith angle of $\rm 37^{\circ}$. An increased antenna spacing of $\rm 50\,m$ led to a small degradation of this resolution to $\rm 4.4\,g\,cm^{-2}$. However, a further increase in the antenna spacing to $\rm 100\,m$ causes a significant degradation in the resolution to a value of $\rm 40.5\,g\,cm^{-2}$, which is approximately a factor of $\rm \sim 13$ times worse than the value obtained for the $\rm 20\,m$ spacing.

\begin{figure}
\includegraphics[width=1\columnwidth]{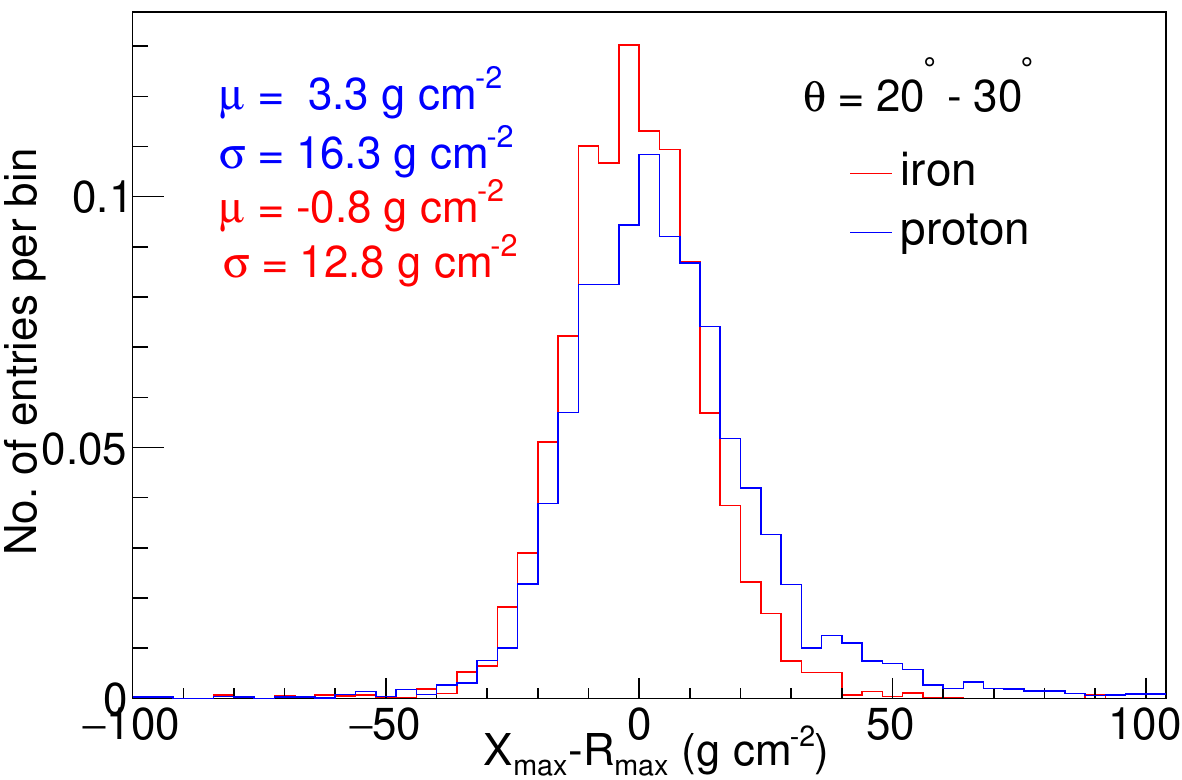}
\caption{$X_\mathrm{max}-R_\mathrm{max}$ distribution (normalized to one) after adding noise for the proton and iron EAS in the energy range of $\rm 10^{17}-10^{17.5}~eV$ and zenith angle $20^{\circ}-30^{\circ}$. The bias $\mu$, obtained as the gaussian peak of the distribution, is $3.3\,\mathrm{g\,cm^{-2}}$ for proton EAS and $-0.8\,\mathrm{g\,cm^{-2}}$ for iron EAS. The spread $\sigma$, defined as the $68\%$ containment region around the peak, is found to be $16.3\,\mathrm{g\,cm^{-2}}$ for proton and $12.8\,\mathrm{g\,cm^{-2}}$ for iron.
}
\label{withnoise}
\end{figure}

In the frequency regime of $\rm 20-80\,MHz$, the galactic and atmospheric noise are  significant in addition to the inbuilt noise from the instruments. In order to estimate the effect of noise on the $X_{\mathrm{max}}$ resolution, we add a Gaussian noise to the simulated radio signals in each antenna following a procedure adopted in Ref. \cite{karastathis2023using}. First, the antenna with the highest amplitude in its dominant polarization is identified, and then the standard deviation of the noise in each antenna is set to a level of $5\%$ of this amplitude. To suppress false triggers, a threshold of $3$ times the standard deviation of the noise is applied. The $X_\mathrm{max}-R_\mathrm{max}$ distributions obtained after adding noise are shown in Figure~\ref{withnoise} for proton and iron EAS of $10^{17}-10^{17.5}$~eV energies in the zenith angle range of $20^{\circ}-30^{\circ}$. The $X_\mathrm{max}$ resolution for proton is found to be $\rm 16.3\,g\,cm^{-2}$ which is $\sim 1.8$ times larger than the resolution without noise. For iron, we obtain $\rm 12.7\,g\,cm^{-2}$ which is  about a factor 2.4 larger than the value without noise. In the higher energy range of $10^{17.5}-10^{18}$\,eV, the resolution after adding noise becomes $\rm 18.8\, g\,cm^{-2}$ for proton and $\rm 14.6\,g\,cm^{-2}$ for iron in the same zenith angle range. The decrease in resolution (larger $X_\mathrm{max}$ values) with energy is due to the shower maximum getting closer to the ground, as discussed in Section-$\rm \Romannum{5}$. 
The effect of the increase in the signal-to-noise ratio with energy is not included because in this analysis the noise level was fixed to a specific fraction of the strongest signal per event. For a more realistic noise model, the $X\mathrm{max}$ resolution is expected to improve with energy. Moreover, there is potential for further improvement by implementing better quality cuts and by fine-tuning the shower reconstruction algorithm. In addition, the antenna response can affect the results. The antenna gain is expected to depend on the polarization of the radio signal at a given antenna position. This can lead to some level of asymmetry in the radio fluence pattern on the ground and affect the reconstructed $X_\mathrm{max}$ value. All of these will be investigated in future work.



The $X_\mathrm{max}$ resolution and the bias obtained with our method (after incorporating noise) appear to be comparable to the values reported for different experiments. Using Monte-Carlo showers, and  following a method that involves fitting the two-dimensional radio lateral distribution, LOFAR reported a resolution of $\rm \sim 17\,g\,cm^{-2}$ and a bias $\rm \lesssim 5\,g\,cm^{-2}$ in the energy range of $\sim \rm 10^{17}-10^{18}\,eV$ \cite{buitink2014method}. A similar method was adopted by the AERA experiment, and found a resolution of $\rm \sim 15-20\,g\,cm^{-2}$ with a bias $\rm \lesssim 10\,g\,cm^{-2}$ in the $\rm 10^{18}-10^{19}\,eV$  range \cite{abdul2024radio}. On the other hand, for the flourescence telescopes at the Pierre Auger observatory and the Telescope Array experiment, a resolution of $\rm \sim 20\,g\,cm^{-2}$ has been reported at energies above $\rm \sim 10^{18}\,eV$. However, at lower energies, these experiments show  poorer resolution: the High Elevation Auger Telescope (HEAT) has a resolution of $\rm \sim 40\, g\,cm^{-2}$ at $\rm 10^{17.2}\,eV$ which increases to $\sim \rm 25\, g\,cm^{-2}$ at $\rm \sim\,10^{18}\,eV$ \cite{bellido2017depth}, and the Telescope Array Low-energy Extension (TALE) reported $\rm \sim 28\,g\,cm^{-2}$ above $\rm\sim 10^{16.5}\,eV$ \cite{fujita2023cosmic}.

\section{Conclusion}
We have presented a new method that has the potential to reconstruct $X_\mathrm{max}$ using radio signals with minimum input from simulation, thereby expected to provide an $X_\mathrm{max}$ value having less uncertainties associated with simulation. The method is based on geometrical reconstruction of the radio emission profile of air shower by backtracking radio signals recorded by an array of antennas at the ground.
The method is also computationally quite efficient, taking only a few CPU seconds for the complete reconstruction of a shower, and require minimal computational resources unlike methods which are based on template or deep learning techniques.


Considering a simple array geometry and an idealized simulation setup without noise, we have demonstrated that the method has the potential to reconstruct $X_\mathrm{max}$ within a resolution of $\rm \sim 14\,g\,cm^{-2}$. We have also shown that this value can be changed when realistic conditions, as in actual CR measurements, are implemented in the simulation. At this stage, the method gives some level of systematic bias in the reconstructed $X_\mathrm{max}$ which should be possible to improve by implementing a better separation of the geomagnetic and charge-excess components, considering the effect of the varying refractive index of air while backtracking the radio signals, reducing the bias in the reconstructed arrival direction of the shower, and using better quality cuts, among others. This will be one of the major focus of our future work. In addition, we will further investigate the effect of noise on the $X_\mathrm{max}$ reconstruction accuracy using realistic models of galactic and electronic noise, and the effect of antenna response as a function of the frequency and arrival direction of the signal, before finally applying the method to real data. The dense array of LOFAR and SKA, and their low-noise environments, will provide an ideal testing ground for our method.

\section{Acknowledgements}
{
We acknowledge funding from the Abu Dhabi Award for Research Excellence (AARE19-224), Khalifa University FSU-2020-13 and the RIG-S-2023-070 grants. We also acknowledge the support from the Czech Science Foundation under the grant number GACR 24-13049S.
}
\clearpage

\appendix
\onecolumngrid
\section{Comparison of the longitudinal profiles of geomagnetic, charge excess, and total emission}
\begin{figure}[h]
\includegraphics[width=1\columnwidth]{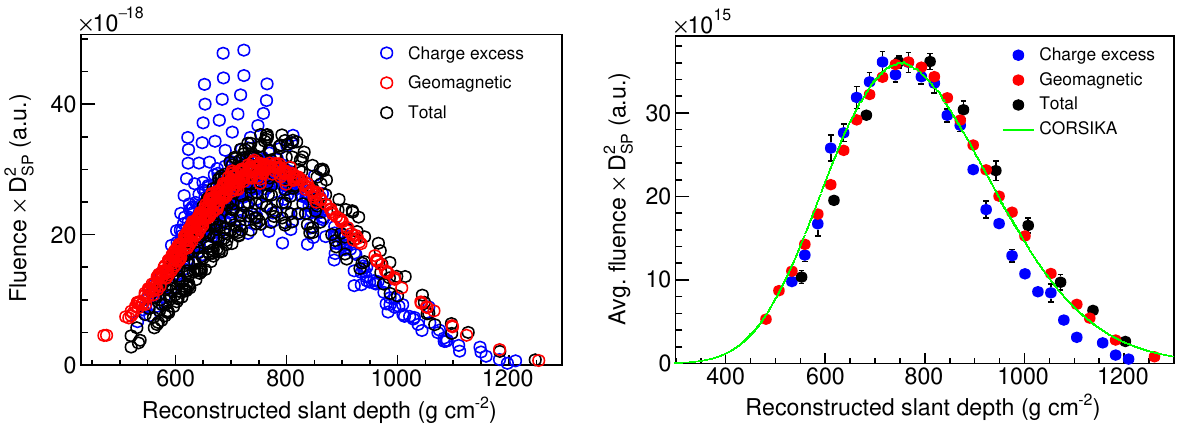}
\caption{ (above) The plot of $\rm fluences\times D^2_{SP}$ as a function of reconstructed slant depth. 
(below) the reconstructed longitudinal profiles (the average of the  $\rm fluences\times D^2_{SP}$, as a function of the slant depth). Here the profiles obtained using the  geomagnetic signal (red), charge excess signal (blue), and the total signal (black) are compared for the EAS shown in Fig.\,\ref{longi}. The  profiles are scaled to match the peak of the true longitudinal profile $\rm N_{e}^{2}$ (green), obtained from CORSIKA. The $X_\mathrm{max}$ of this EAS is $\rm 753.5\,g\,cm^{-2}$. The arrival direction $\rm(\uptheta,\upphi)$, and the energy of this proton EAS are $\rm (37.8^{\circ},154.7^{\circ})$, and $\rm 2.5\times10^{8}\,eV$ respectively. A Gaisser-Hillas fit to the peak of each  longitudinal profile yields the reconstructed $R_\mathrm{max}$ to be $\rm 776.9\,\pm\,9.5\,g\,cm^{-2}$, $\rm 764.1\,\pm\,1.7\,g\,cm^{-2}$, and $\rm 749.6\,\pm\,3.5\,g\,cm^{-2}$ for the total, geomagnetic, and charge excess emission respectively. The charge excess emission is scaled by a factor of $\sim 53$ in the plot.
}
\label{geoexecomp}
\end{figure}

In this section, the reconstructed longitudinal profiles obtained from the total, geomagnetic and the charge excess signals are compared. As shown in the left panel of Figure~\ref{geoexecomp}, a large scatter is seen for the total emission (black points), owing to the asymmetry of the radio footprint pattern. Although the scatter appears larger for the charge excess, this is an artificial artifact, as the graph for the charge excess emission has been scaled along the Y-axis. By fitting the reconstructed radio emission profile obtained using the total emission with the Gaisser-Hillas function as shown in the right panel, we obtain an $R_\mathrm{max}$ value of $\rm 776.9\,g\,cm^{-2}\pm 9.5$, which is higher than that obtained from the geomagnetic signal by $\rm 12.8\,\,g\,cm^{-2}$. The $R_\mathrm{max}$ value for the charge excess emission is smaller than the geomagnetic emission by $ \rm 14.5\,g\,cm^{-2}$. 
\vspace{1mm}

\label{app1}
\section{The longitudinal profiles of EAS}
\begin{figure}
\centering
\includegraphics[width=1\textwidth]{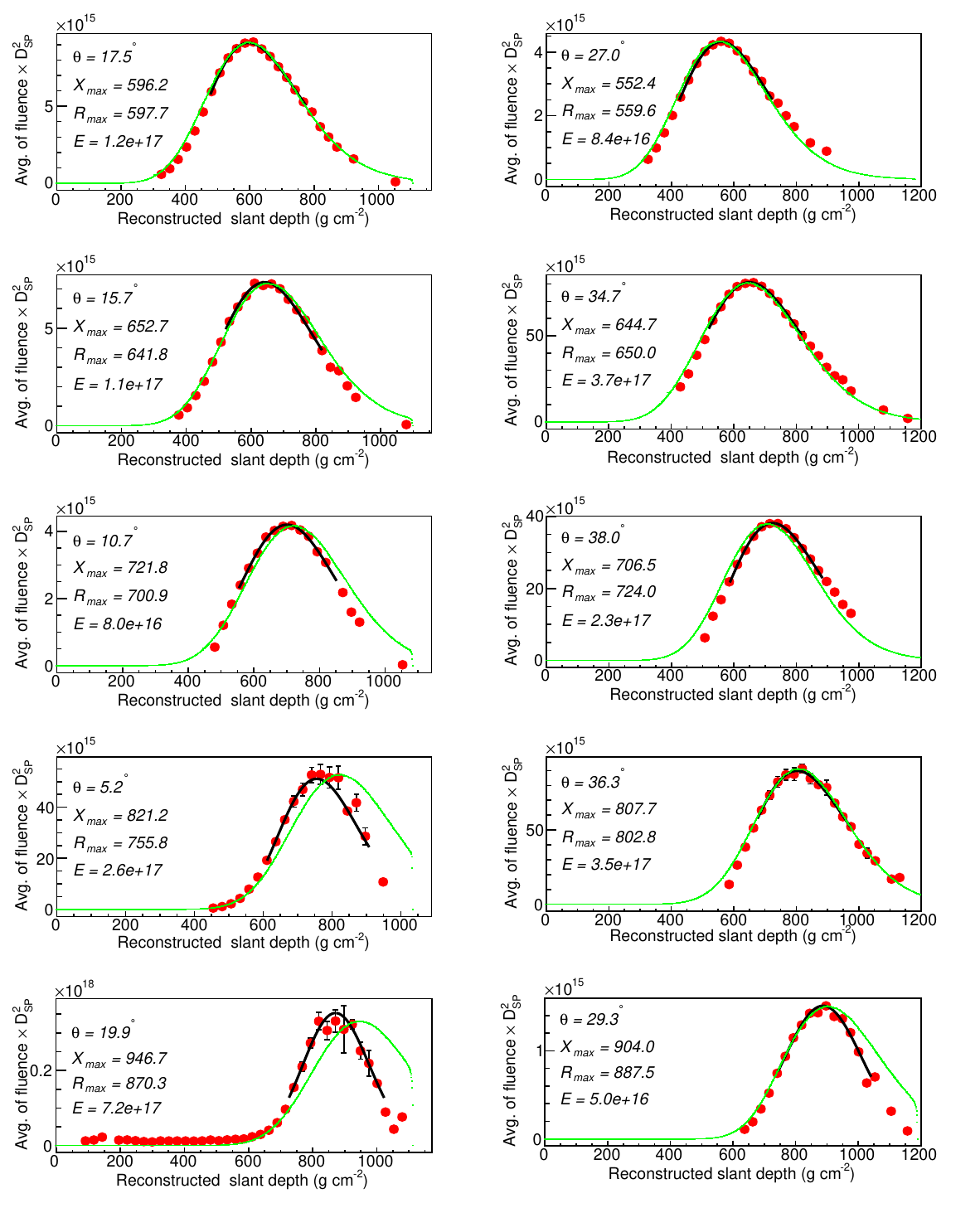}
\caption{Reconstructed radio emission profiles of proton EAS obtained using the geomagnetic signal (red) compared with the shower longitudinal profile $N_\mathrm{e}^{2}$ obtained from CORSIKA (green). The values of $R_\mathrm{max}$ for the reconstructed profile and $X_\mathrm{max}$ for the shower profile are given in units of $\rm g\,cm^{-2}$. The zenith angle $\rm \uptheta$ (in degrees) and the energy $E$ (in eV) of the EAS are also shown. The $R_\mathrm{max}$ value is obtained by fitting the Gaisser-Hillas function around the peak of the distribution (black line). The fourth and the fifth plots on the left column are EAS with $d_{max}\lesssim 2\,km$, which are not so well reconstructed.}
\label{longiprof_20}
\end{figure}

\begin{figure}
\centering
\includegraphics[width=1\textwidth]{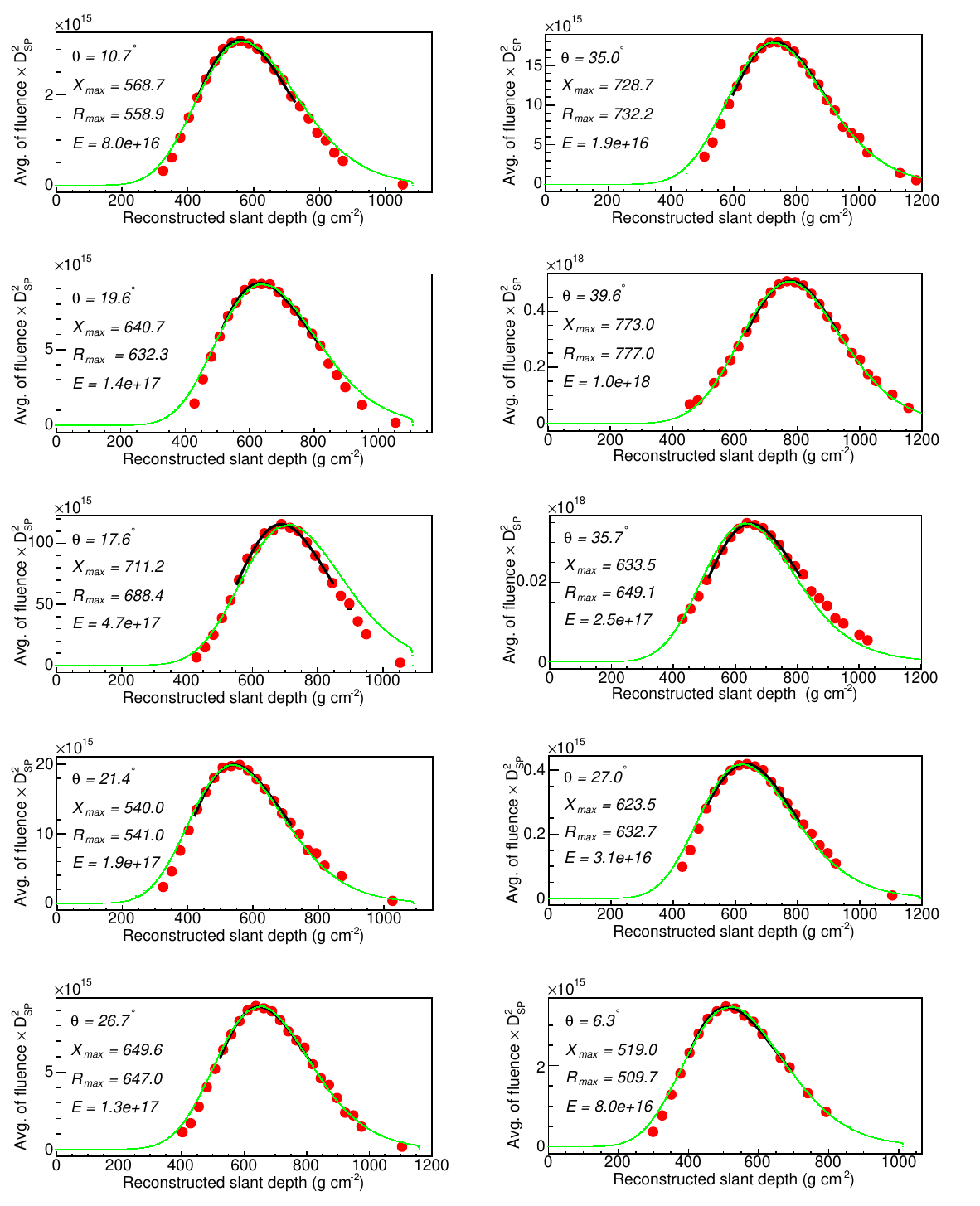}
\caption{Reconstructed radio emission profiles of iron EAS obtained using the geomagnetic signal (red) compared with the shower longitudinal profile $N_\mathrm{e}^{2}$ obtained from CORSIKA (green). The values of $R_\mathrm{max}$ for the reconstructed profile and $X_\mathrm{max}$ for the shower profile are given in units of $\rm g\,cm^{-2}$. The zenith angle $\rm \uptheta$ (in degrees) and the energy $E$ (in eV) of the EAS are also shown. The $R_\mathrm{max}$ value is obtained by fitting the Gaisser-Hillas function around the peak of the distribution (black line).}
\label{longiprof_20_iron}
\end{figure}

The radio emission profiles of proton EAS that have been reconstructed using the new method are shown in Figure~\ref{longiprof_20}. For most of the EAS shown, the reconstructed radio profiles are found to agree quite well with the shower longitudinal profile (distribution of $N_\mathrm{e}^{2}$) obtained from CORSIKA. However, EAS with $X_\mathrm{max}$ close to the ground ($\rm \lesssim 2\,km$) are not well reconstructed. For a vertical EAS, a distance  of $\rm 2\,km$ above the ground translates to a slant depth  of $\rm \sim 800\,g\,cm^{-2}$, whereas for an EAS arriving at a zenith angle $\rm \uptheta$ of $\rm 40^{\circ}$, the same distance translates to a slant depth of  $\rm\sim 1100\,g\,cm^{-2}$. The reconstructed radio profiles of iron EAS are shown in Figure~\ref{longiprof_20_iron}.
\vspace{0.3mm}

\label{app2}

\clearpage

\section{Distribution of $(X_{\rm max}-R_{\rm max})$ for the $\rm 10^{17.5}\,eV$-$\rm 10^{18}\,eV$ range}
\label{appendixC}
\begin{figure}[H]
\centering
\includegraphics[width=1\textwidth]{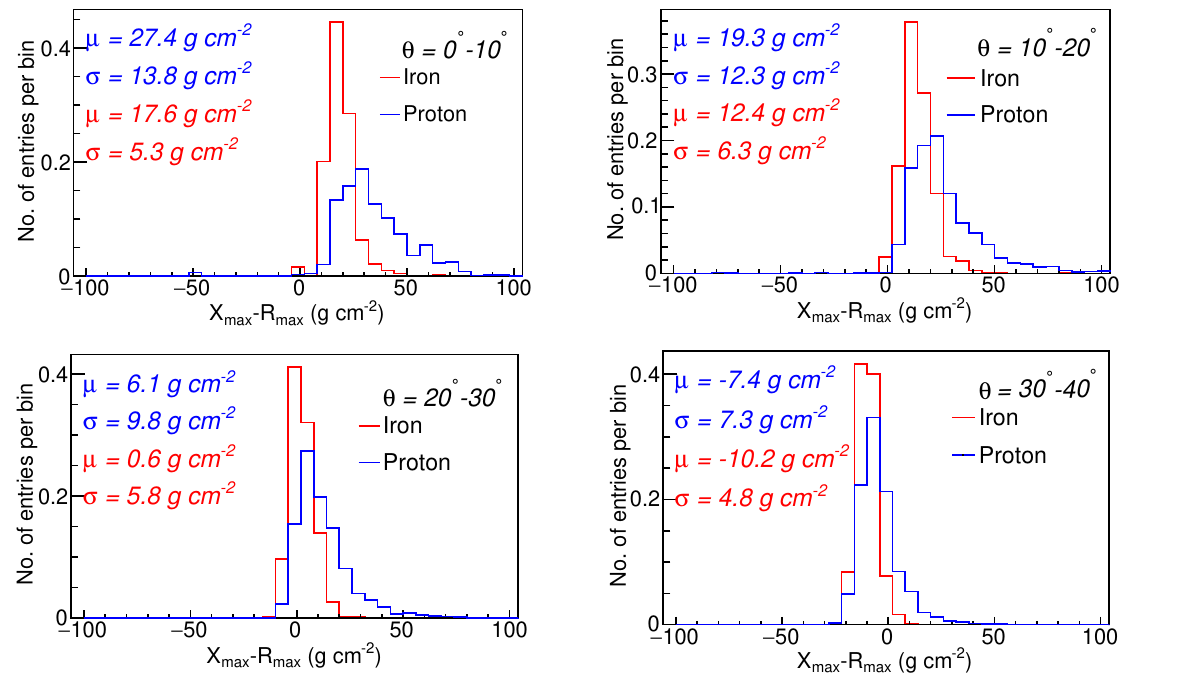}
\caption{$X_\mathrm{max}-R_\mathrm{max}$ distribution (normalized to one) for the proton (blue) and iron EAS (red) in the energy range of $\rm 10^{17.5}-10^{18}\,eV$ for four different zenith bins: $\rm 0^{\circ}-10^{\circ}$, $\rm 10^{\circ}-20^{\circ}$, $\rm 20^{\circ}-30^{\circ}$ and $\rm 30^{\circ}-40^{\circ}$. The distributions are obtained without including noise and electronic response in the analysis. The bias $\mu$ are determined by fitting a Gaussian function around the peak of the distribution, and the spread $\sigma$ represent the $68\%$ containment region from the peak. Note that the $\rm 68\%$ containment region is determined by using a finer bin width of $\rm 0.5\,g\,cm^{-2}$.
}
\label{xmaxdiff2}
\end{figure}

Figure \ref{xmaxdiff2} shows the $(X_{\rm max}-R_{\rm max})$ distribution of the proton and iron showers for the energy range of $\rm 10^{17.5}-\rm 10^{18}\,eV$ in four different zenith angle bins between $\rm 0^{\circ}$ and $\rm 40^{\circ}$. 

\clearpage
\bibliographystyle{apsrev4-1}
\bibliography{references.bib}
\end{document}